\documentclass[11pt,letterpaper]{article}
\usepackage{arxiv}

\usepackage[pdftex]{graphicx}
\usepackage{graphicx}
\usepackage{rotate}
\usepackage{wrapfig}
\usepackage{caption}
\usepackage[T1]{fontenc}
\usepackage[english]{babel}
\usepackage{times}
\usepackage{amsmath,amstext,amsbsy,amsopn,amsthm,amsfonts,amssymb}
\usepackage{mathtools}
\usepackage{url}
\usepackage{tabularx,booktabs}
\newcolumntype{C}{>{\centering\arraybackslash}X}
\usepackage{booktabs,array,colortbl}
\usepackage{lastpage}
\usepackage{multirow}
\usepackage{algorithm}
\usepackage{balance}
\usepackage{placeins}
\usepackage{array}
\usepackage{blindtext}
\usepackage{subfigure}
\usepackage{caption}
\usepackage{capt-of}
\usepackage{booktabs}
\usepackage[small]{titlesec}
\usepackage{paralist}
\usepackage{verbatim} 
\usepackage[usenames,dvipsnames,svgnames]{xcolor}
\usepackage{algorithmic}
\usepackage{fancyhdr}
\usepackage{wrapfig}
\providecommand{\keywords}[1]{\textbf{\textit{Keywords---}} #1}
\begin{document}

\title{iGLU 2.0: A New Non-Invasive, Accurate Serum Glucometer for Smart Healthcare}



\author{
\begin{tabular}{cccc}
Prateek Jain & Amit M. Joshi & Navneet Agrawal & Saraju P. Mohanty \\
ECE Department & ECE Department & Diabetologist & CSE Department\\
MNIT, Jaipur, India. & 	MNIT, Jaipur, India. & Diabetes Center, India & UNT, USA. \\
prtk.ieju@gmail.com &amjoshi.ece@mnit.ac.in & navdotc@gmail.com & saraju.mohanty@unt.edu
\end{tabular}	
}

\maketitle

\cfoot{Page -- \thepage-of-\pageref{LastPage}}

\begin{abstract}
\textit{To best of authors knowledge, this article presents the first-ever non-invasive glucometer that takes into account serum glucose for high accuracy.}
In case of blood glucose measurement, serum glucose value has always been considered precise blood glucose value during prandial modes. Serum glucose can be measured in laboratory and more stable glucose level compare to capillary glucose. However, this invasive approach is not convenient for frequent measurement. Sometimes, Conventional invasive blood glucose measurement may be responsible for cause of trauma and chance of blood related infections. To overcome this issue, in the current paper, we propose a novel Internet-of-Medical (IoMT) enabled glucometer for non-invasive precise serum glucose measurement. In this work, a near-infrared (NIR) spectroscopic technique has been used for glucose measurement. The novel device called iGLU 2.0 is based on optical detection and precise machine learning (ML) regression models. 
The optimal multiple polynomial regression and deep neural network models have been presented to analyze the precise measurement. The glucose values of serum are saved on cloud through open IoT platform for endocrinologist at remote location. To validate iGLU 2.0, Mean Absolute Relative Difference (mARD) and Average Error (AvgE) are obtained 6.07\% and 6.09\%, respectively from predicted blood glucose values for capillary glucose. For serum glucose, mARD and AvgE are found 4.86\%  and 4.88\%, respectively. These results represent that proposed non-invasive glucose measurement device is more precise for serum glucose compared to capillary glucose. 
\end{abstract}



\section{Introduction}

Remote connectivity of doctors and patients is the key solution in providing better and advanced medical facilities to the patients \cite{Zhu_MCE_2019-Sep, Mohanty_CEM_2016-Jul}. As the healthcare technologies are being advanced, there is an increment and consciousness of consumers for their health. In particular scenario, the demand for remote healthcare is more promoted than ever. Present Internet-of-Medical-Things (IoMT) solution for smart healthcare encourages hospitals to ameliorate the care quality with focusing on overall expenses reduction. The basic requirement of our Smart Healthcare solution comprises of sensors that collect patient data, cloud to store, analyze and process the data, web services and mobile applications for patients and doctors. In a big picture IoMT enabled healthcare Cyber-Physical System (CPS) makes smart healthcare possible \cite{Mohanty_IEEE-MetroCon-2019_Invited-Talk}.

\begin{figure}[htbp]
	\centering
	\includegraphics[width=0.85\textwidth]{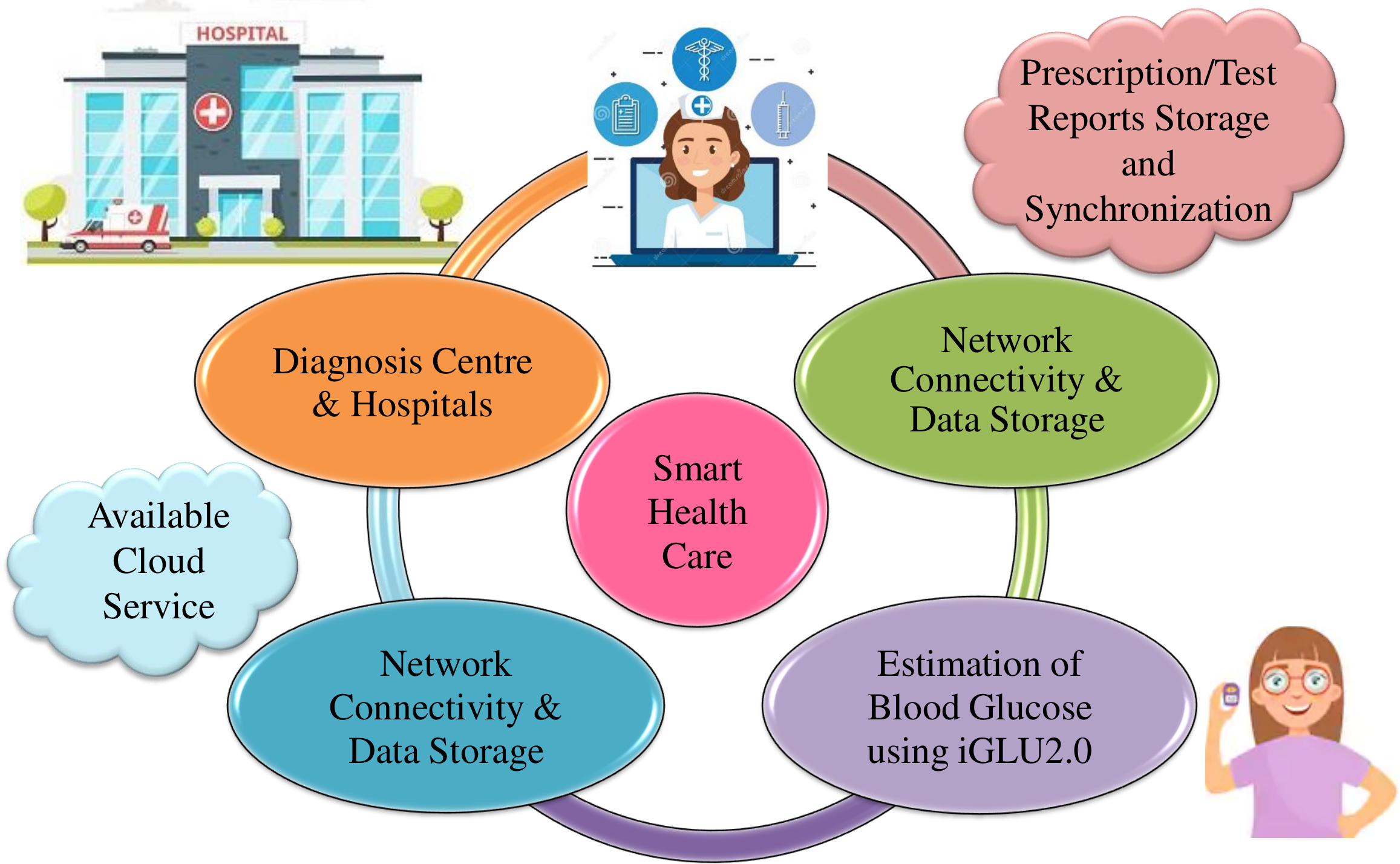}
	\caption{Blood glucose diagnosis in smart health care system.}
	\label{smart}
\end{figure}

The non-invasive, precise and rapid diagnosis of diabetes is of great demand worldwide due to low-cost, painless, and easy of usage \cite{Jain_IEEE-MCE_2020-Jan_iGLU1, Jain_arXiv_2019-Nov30-1911-04471_iGLU1, IEEE_Std_11073-10425-2017, Wang_TBCS_2017-Oct}. Instant non-invasive serum glucose measurement is not being possible at present to overcome the cause of trauma. Hence, the instant measurement of serum glucose with continuous monitoring is being recent challenge in the smart healthcare system. Proposed serum glucose measurement device is a handheld, IoMT enabled end-device which provides rapid and continuous monitoring. It interacts with an endocrinologist to the remote located diabetic patients. The process flow of blood glucose diagnosis in smart health care system is shown in Figure \ref{smart}, which represents the smart healthcare system for diabetic patients at remote  location.

\begin{figure}[htbp]
	\centering
	\includegraphics[width=0.85\textwidth]{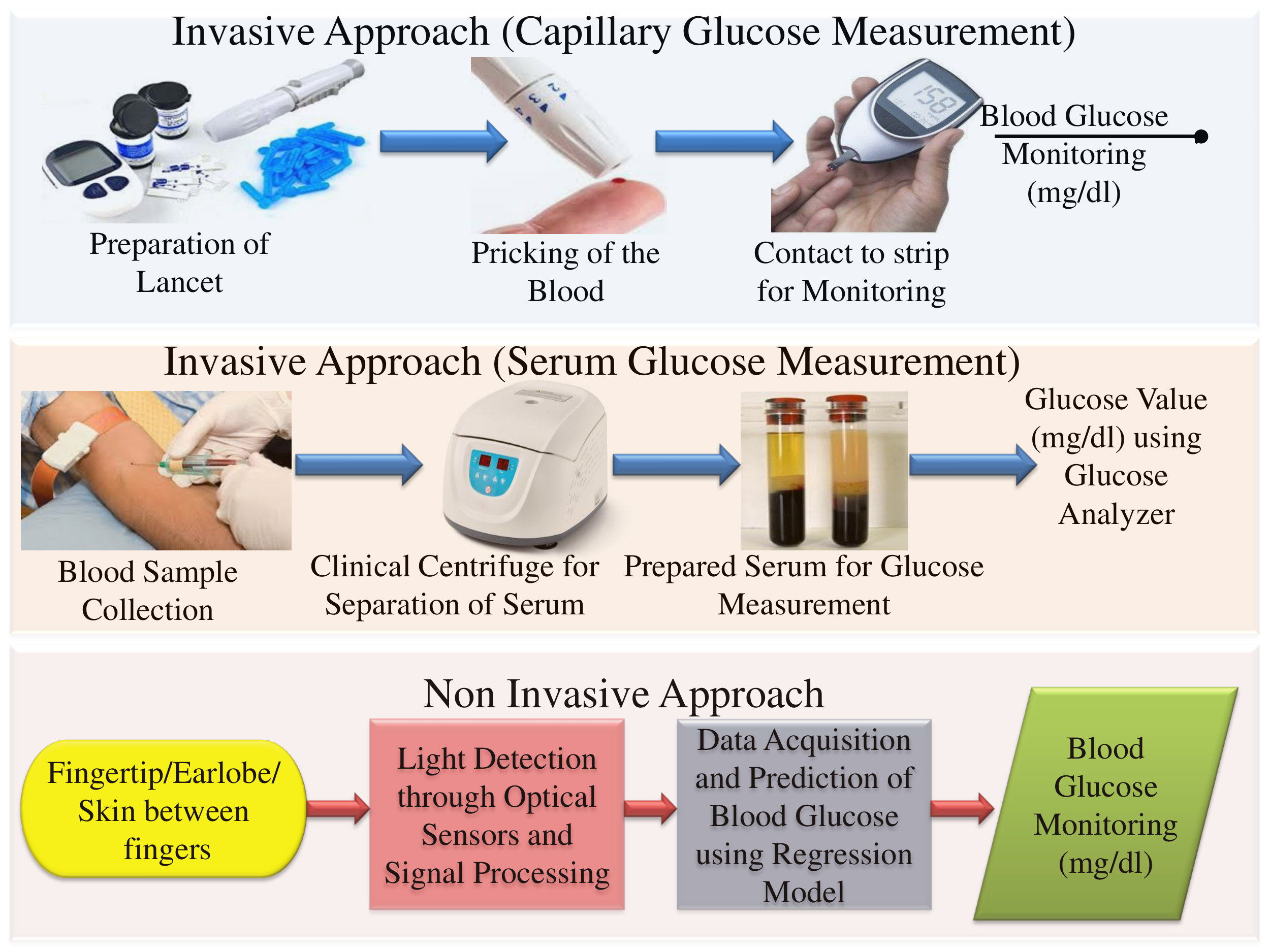}
	\caption{Processing steps of non invasive blood glucose measurement device.}
	\label{theme}
\end{figure}

In proposed non-invasive serum glucose measurement, optical detection approach is involved. The collected data is processed for acquisition and glucose is predicted by the regression model. The process flow is shown in Figure \ref{theme}. 
According to this process, patients measure their serum glucose without pricking blood and store directly to the cloud where nearby endocrinologist can analyze the serum glucose data of particular patient. The prescription would also be sent by endocrinologist to particular patient for further treatment. Traditionally, blood glucose is being measured in the form of serum and capillary glucose of diabetic patients. The serum glucose test is a laboratory test where serum (centrifuged blood) is extracted from blood to measure the glucose level, whereas capillary glucose is being measured by commercially available glucometer. In this process, a drop of blood is pricked through lancet from finger and blood drop is taken on strip which has to be connected to the glucometer. For accurate blood glucose value, diabetic patients are always recommended for the serum glucose test. But, the whole process takes 1-2 hours for serum glucose value. Hence, it is necessary to develop the device which can measure the serum glucose without puncturing the finger instantly. People will be more conscious for blood glucose continuous monitoring. Such a device is advisable for smart healthcare. 
Designing a non-invasive serum glucose monitoring device comprises of spectroscopy techniques. Spectroscopy is the interaction between optical radiation and any molecules.
This technique is used to measure the resultant light after interaction with molecules at the specific resonance frequency. 

The rest of the paper is composed in the following manner. The prior related work is described in Section \ref{Sec:Prior-Works}. Novel contribution of this paper is represented in Section \ref{Sec:Novel}. Section \ref{Sec:Calibration} discusses analytical modelling and calibration details of the sensor for serum and capillary glucose. The proposed device is presented with mechanism of serum glucose detection in Section \ref{Sec:Proposed-Work}. Experiments and analytical modelling with error analysis have been elaborated in Section \ref{Sec:Experimental-Results}.

\section{The State-of-Art in Glucose Measurement and its Advancement through the Current Paper}
\label{Sec:Prior-Works}

\subsection{Prior Research on Glucose Level Monitoring}

\begin{figure}[htbp]
	\centering
	\includegraphics[width=0.995\textwidth]{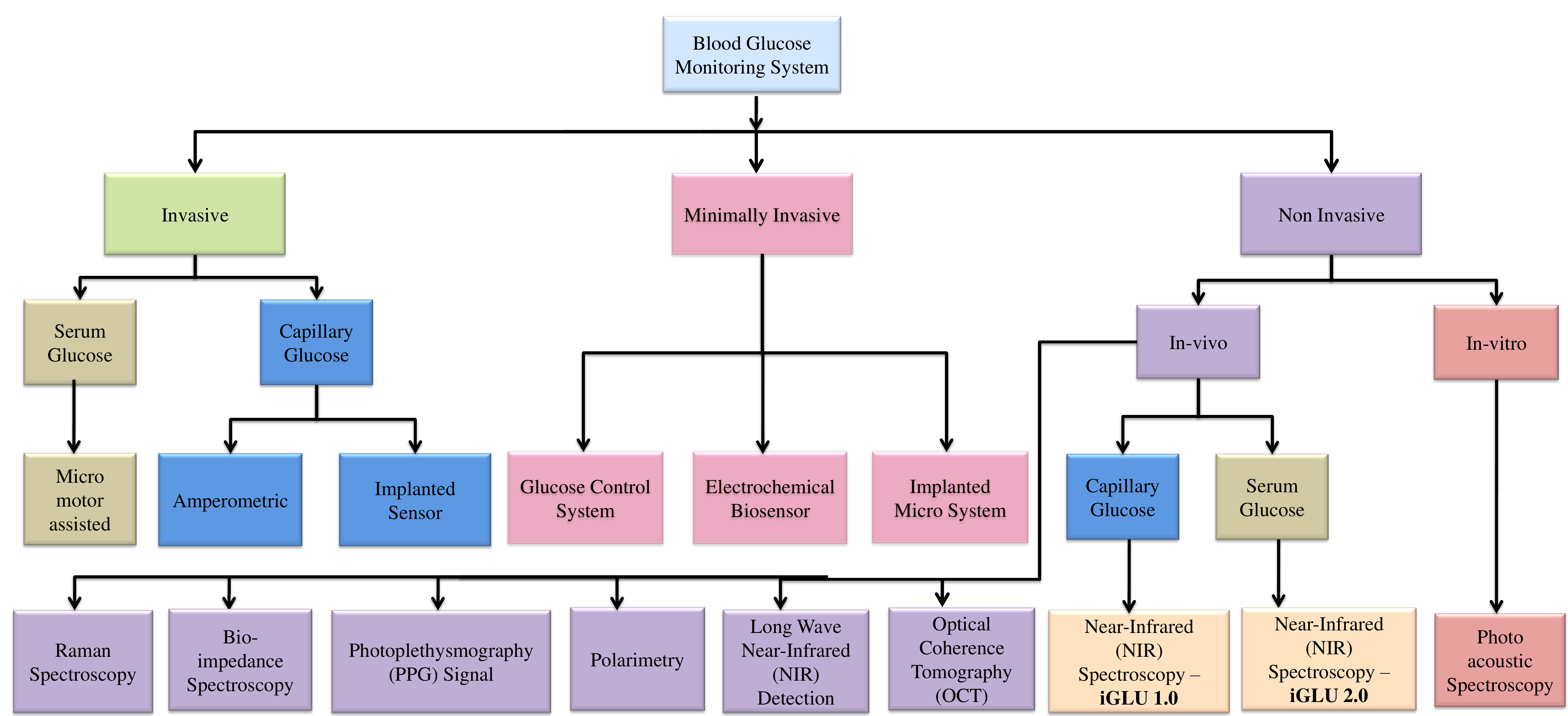}
\caption{An overview of the related prior works.}
	\label{priorwork}
\end{figure}

Research on glucose level monitoring is being undertaken for a long time and also ongoing. Related prior research work can be classified as invasive, minimally invasive, and non-invasive blood glucose monitoring as represented in Figure \ref{priorwork}.

Various invasive and non-invasive approaches have been explored for glucose measurement \cite{Beach2005}. Demitri, et al. represented the photometric approach for glucose measurement using small blood volumes \cite{demitri2017measuring}. To mitigate the issue of frequent pricking blood, one-time wearable sensor is proposed in terms of a semi-invasive approach. This wearable sensor has advantages of rapid diagnosis and continuous monitoring. But, this has to be changed after a certain period. Hence, this will be irritating and may be the cause of trauma. 
Wang, et al. represented wearable minimally invasive microsystem for glucose monitoring \cite{Wang_TBCS_2017-Oct}. This wearable microsystem is a neither painless nor cost-effective solution. Acciaroli, et al. discussed the new method to reduce the frequency of calibration of minimally invasive Dexcom sensor \cite{acciaroli2018reduction}. 
To make the painless system, Pai, et al. designed the prototype setup of photoacoustic spectroscopy for non-invasive glucose measurement \cite{Pai2018}. But, implementation of corresponding components yield the setup costly and also occupies large area. However, the solution will not be portable. Yin, et al. proposed a DiabDeep framework that merge wearable sensors and neural networks for pervasive diagnosis of diabetes. They achieved classification accuracy to classify the healthy and diabetic patients. They considered various environmental and physiological parameters to justify the diabetic patients. However, accuracy has not been identified in terms of blood glucose measurement \cite{Yin_2019}. The work did not represent any kind of error analysis. Prasad, et. al. presented nanoparticles on alkali anodized steel (AS) electrode for glucose measurement through saliva \cite{8347021}. Singh, et al. represented optical biosensor for glucose measurement using saliva \cite{8727488}. Glucose measurement has also been done using IMPS spectroscopy through the skin \cite{Dai2009, Song2015}. Electrical properties of skin, sweat and saliva vary according to person. So, this approach will not be reliable for glucose measurement. Non-invasive glucose measurement approach through retina has also been represented for precise glucose detection \cite{de2016optical}. Such non-invasive approach is not desirable for frequent monitoring \cite{pirnstill2012vivo}. Raman spectroscopy has been preented for precise glucose measurement \cite{shih2015noninvasive}. The implemented setup occupied a large area and will not be portable.

\subsection{Commercial Products on Glucose Level Monitoring}

Various products such as glucometer from Labiotech \cite{Fernandez2020}, GlucoTrack\textsuperscript{\textregistered}, and so on have limitations in terms of precision. The cost of these products is also high which varies in the range of 300-400 USD. Therefore, the cost effective device for precise non-invasive blood glucose measurement is presented. The Medical Training Initiative (MTI) is developing a non-invasive glucose monitoring device glucowise\textsuperscript{\texttrademark}. 2M Engineering is also actively developing solution for non-invasive glucose monitoring \cite{2020me}. The device is based on Raman scattering spectroscopy technique with the use of laser technology. However, this is not cost effective solution. These devices are not available in the market for medication. In this way Freestyle Libre sensor, SugarBEAT from Nemaura medical are available in the market for medication \cite{sugarbeat}. These products are adhesive and skin-patch which are regularly disposable. These devices support the continuous glucose monitoring. These are minimally invasive devices. Omelon B-2 is one of the non-invasive stripless devices for continuous glucose monitoring which is available in the market for medication \cite{omelon}. The device is bulky and is not wearable device for continuous glucose level monitoring and the precision in measurement has also not been achieved according to the users remarks.

\subsection{Why NIR over other Non-Invasive Approaches?}

\begin{figure}[htbp]
	\centering
	\includegraphics[width=0.85\textwidth]{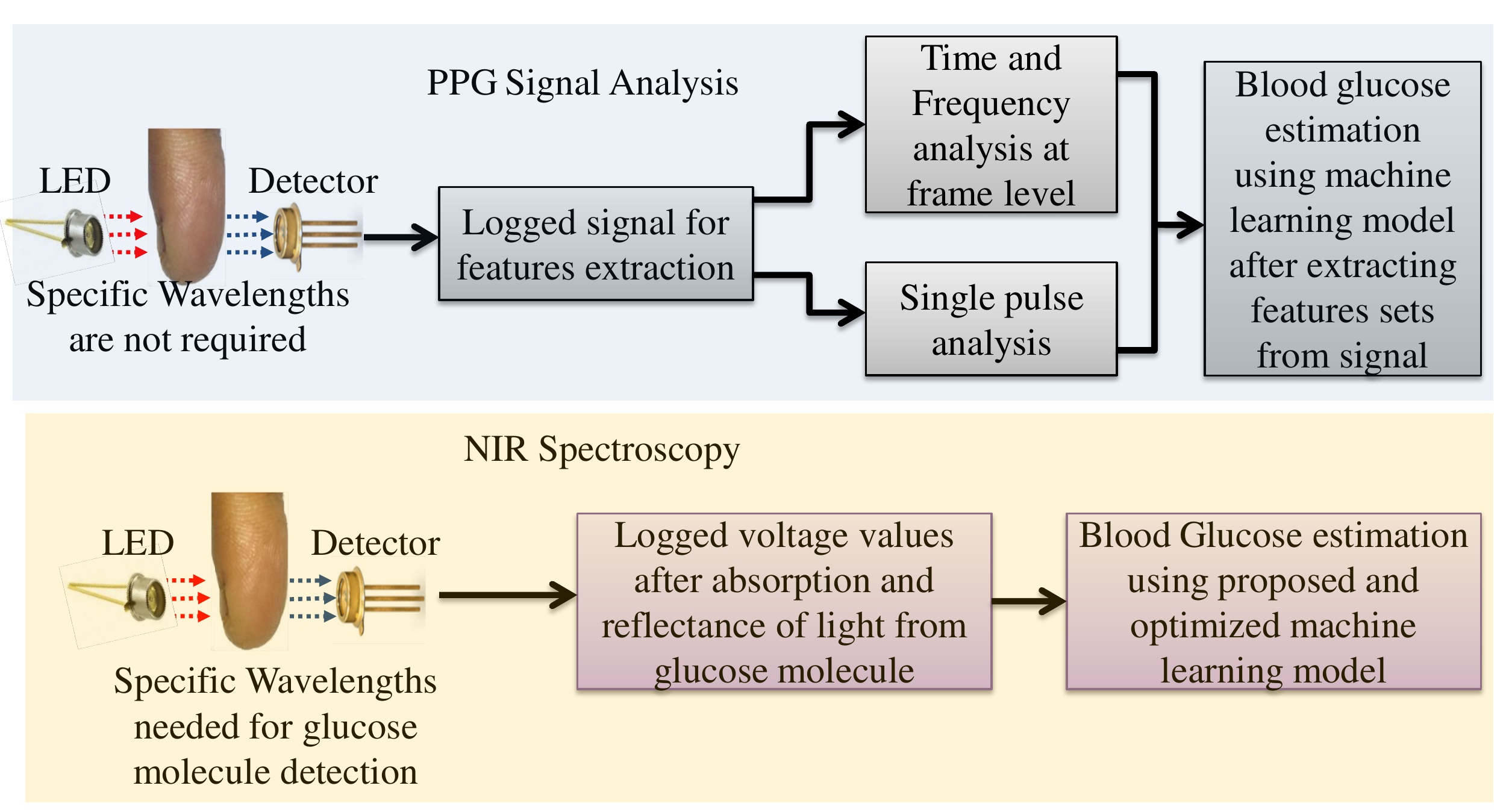}
	\caption{Mechanism of Serum and Capillary Glucose Measurement.}
	\label{ppgvsnirs}
\end{figure}

Glucose measurement has been done using various non-invasive approaches such as impedance spectroscopy, NIR light spectroscopy, PPG signal analysis and so on. But, apart from optical detection, other techniques have constraints in terms of precise measurement. Desirable accuracy is not possible from measurement through sweat and saliva as properties always vary for each person. PPG signal analysis is based on extracted features of logged signal which is not based on principle of glucose molecular detection \cite{monte2011non, habbu2019estimation}. The methodology of glucose measurement through PPG signal and NIRS are represented in Figure \ref{ppgvsnirs}.

To overcome these limitations, Sharma, et al. also discussed about optical detection using long NIR wave which is not capable to detect the glucose molecules beneath the skin as it has shallow penetration \cite{sharma2013efficient}. Therefore, small NIR wave has been chosen for real-time glucose detection \cite{Ali2017, haxha2016optical}.

\subsection{Why Serum Glucose over Capillary Glucose?}

In the case of invasive approaches, serum glucose and capillary glucose level are being analyzed for precise blood glucose values. Capillary glucose can be estimated instantly but serum glucose couldn't be identified instantly due to certain processes. However, it is clinically observed that the capillary glucose level is always higher than serum glucose which is not being considered actual blood glucose ever. Hence, there is a trade-off between both approaches. But, serum glucose has always been recommended for diagnosis as an accuracy point of view. Therefore, serum glucose is always being reliable compared to capillary glucose for medication.

\begin{figure}[htbp]
	\centering
	\includegraphics[width=0.85\textwidth]{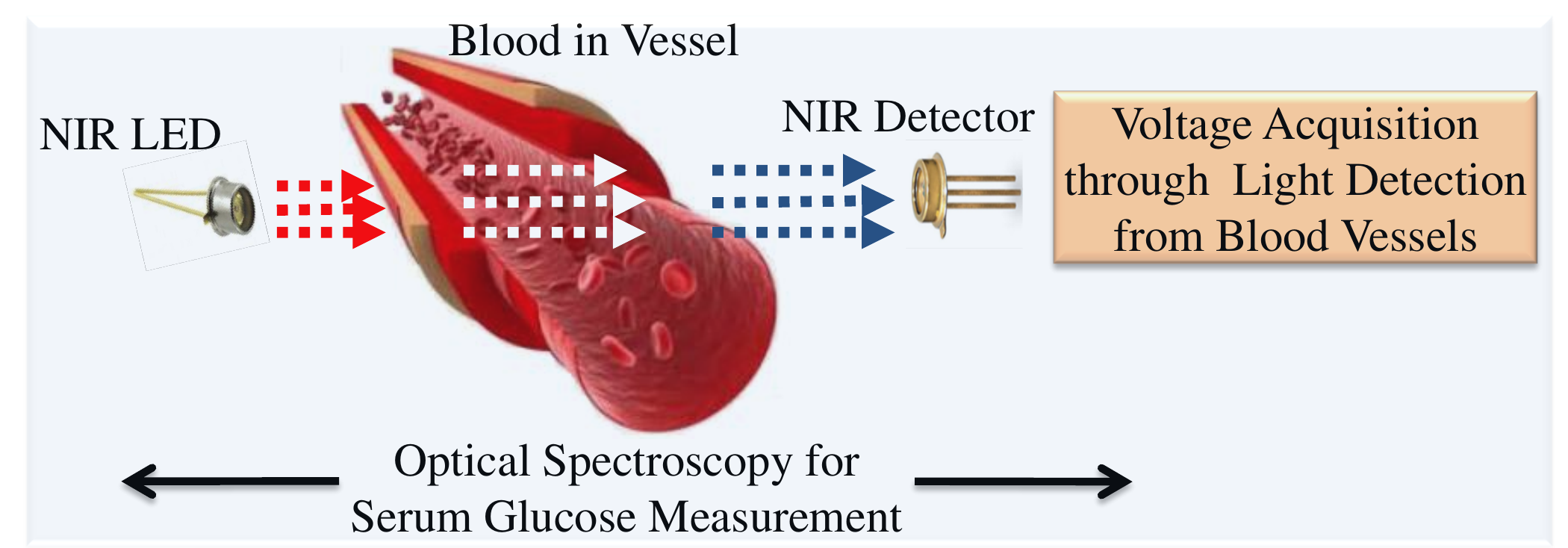}
	\caption{NIR Spectroscopy Mechanism of Serum Glucose Measurement.}
	\label{serumvscapillaryglucose}
\end{figure}

The serum glucose measurement is explored through NIR light as presented in Figure \ref{serumvscapillaryglucose}. The incident light is passed through blood vessels and blood glucose molecules are detected in flowing concentration. During capillary glucose measurement, the drop of blood is pricked from epidermis layer of skin. The precise value of glucose may not be possible always through capillary blood glucose measurement. Therefore, serum glucose measurement is advisable through diabetologist generally for precise measurement.

\subsection{Advancement over our Previous work iGLU}

The works related to non-invasive blood glucose measurement have been proposed using multiple techniques. Non-invasive detection reduces the chance of being blood-related infections during measurement. But, these measurement techniques have some limitations. Hence, the device iGLU 1.0 has been represented for non-invasive capillary glucose measurement \cite{Jain_IEEE-MCE_2020-Jan_iGLU1}. Serum glucose has always been considered more precise compared to capillary blood glucose as per medical aspects. Therefore, it is required to design a portable device for non-invasive serum glucose measurement which would have better accuracy. In current paper, iGLU 2.0 is proposed for serum glucose monitoring using short-wave NIR spectroscopy technique with calibration and validation \cite{jain2019precise}.

\section{Novel Contributions of the Current Paper to the State-of-Art}
\label{Sec:Novel}

\textbf{Research Questions Addressed in the current Paper}: For serum glucose measurement, it is necessary to prick the blood and it takes more time to measure the glucose. The well equipped laboratory is required for processes to extract the serum and storage at specific (frozen) temperature. Hence, the laboratory needs a large space with frozen facility of serum. Hence, this process is neither being a cost-effective nor time-saving solution in terms of instant diagnosis. The questions which are addressed in the current paper for the advancement of smart healthcare: (1) How can we have a device that can measure serum glucose without pricking the blood at the user and stores the data in cloud for future use by the patient and healthcare providers? (2) Can we have a device that can provide the facility of continuous serum glucose monitoring at the user-end so that it can function even if the Internet connectivity is not available all the time? (3) Can we have a device that can measure automatically serum glucose frequently all the time? (4) Can we have a device that can measure the serum glucose of all types of patients precisely?

\textbf{Challenges in Addressing the Research Problem and Questions}:
The solution of various challenges for the research questions to obtain precise serum glucose measurement include the following:
(1) Analyzing and validation of specific wavelengths for glucose detection are required for precise measurement. (2) Optimization in circuit level design of data acquisition module is required to improve the performance of proposed device for CGM. (3) Acquiring of voltages and serially transfer for data logging will be possible with synchronization for frequent measurement. (4) Analysis of optimized regression model with calibration and validation using healthy and diabetic samples to measure serum glucose of all patients precisely.    

\begin{figure}[htbp]
	\centering
	\includegraphics[width=0.85\textwidth]{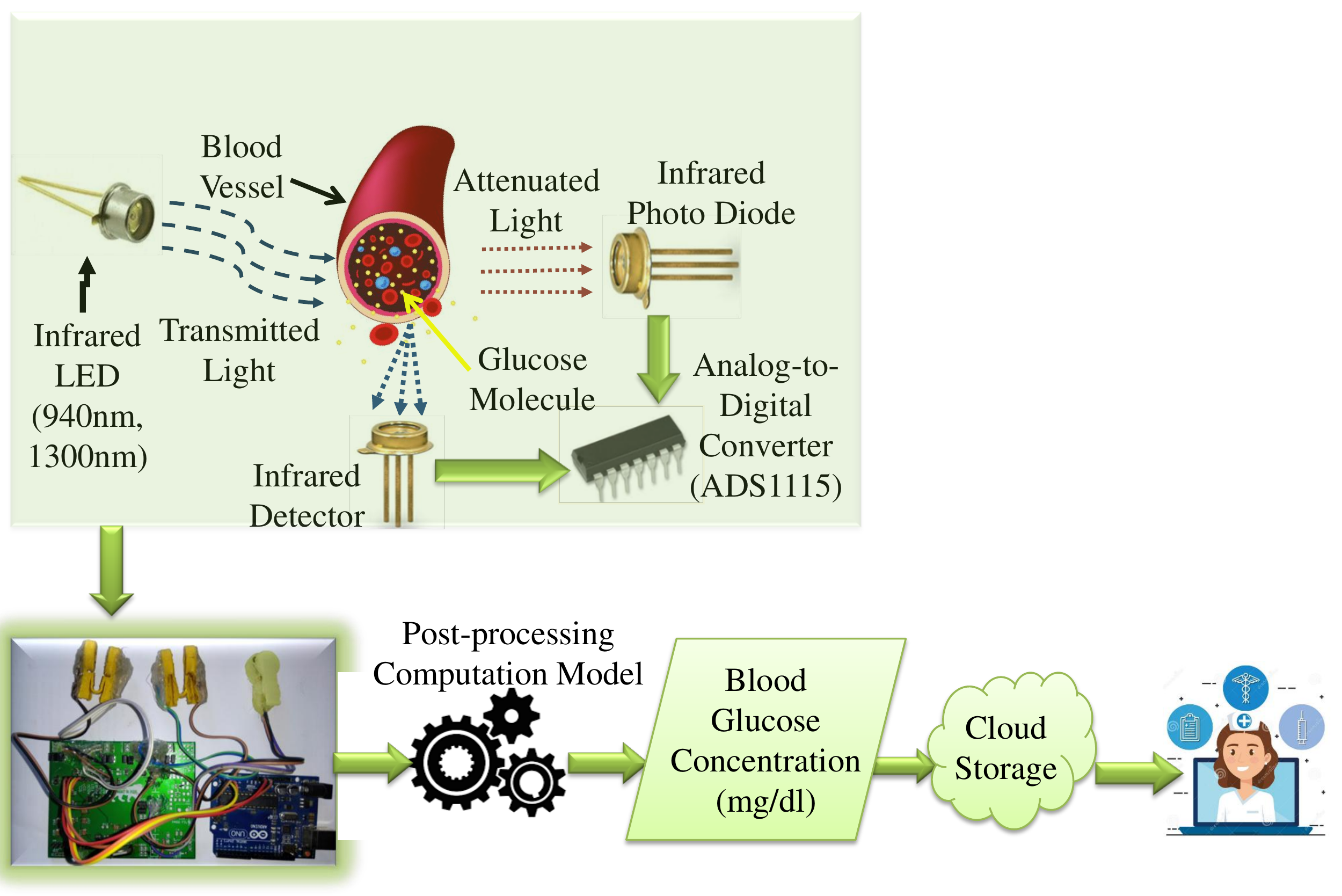}
	\caption{Top level representation of proposed device (iGLU 2.0).}
	\label{FIG:Proposed_Glucometer_Overview}
\end{figure}

\textbf{Proposed Solution of this Paper:} To address the above research questions, we proposed an edge-device called iGLU 2.0 in the current paper that measures serum glucose of patient and stores the data at the cloud. We have proposed the non-invasive serum glucose measurement device which is a comparatively precise and low-cost solution. The proposed device is also integrated with IoMT framework for data storage on the cloud. The risk factor of blood-related diseases has been mitigated through proposed device. The device can be used to measure the serum glucose of every person at anytime. iGLU 2.0 is an advancement in the non-invasive device era. This device measures the serum glucose instantly with continuous monitoring. Flow of proposed work is represented in Figure \ref{FIG:Proposed_Glucometer_Overview}.

\textbf{Novel contribution in current paper} include the following: 
\begin{enumerate}
	\item
	A precise non-invasive serum glucometer called iGLU 2.0 using absorption and reflectance of light with two short NIR waves using specific wavelengths (940 nm and 1300 nm) has been introduced.
	\item
	An optimized kernel based regression model for serum glucose sensor calibration has been presented.
	\item
	A serum glucose monitoring device has been formed with optimized circuital biasing of sensors for calibration using precise and optimized regression model to validate and test the subjects.   
	\item 
	With the active support from diabetic center and hospital, real-life experimental validation has been done directly from human blood.
	\item
	The proposed device has been integrated in IoMT framework for data (serum glucose values) storage, patient monitoring and treatment on proper time with cloud access by both the patient and doctor.
\end{enumerate}

\section{Machine Learning (ML) Models for Blood Glucose Level Calculation from the NIR Signal}
\label{Sec:Calibration}

Optimized computation model is analyzed after calibration of regression models (RM) for serum and capillary glucose estimation. In current work, the detector's outputs from three channels are logged as input vectors for prediction of glucose. The calibrated models are used to predict the blood glucose concentrations for validation. The logged data from each sample is essential to convert in predicted glucose values. Hence, it is required to calibrate an appropriate kernel for precise measurement. 113 samples of capillary glucose and 74 samples of serum glucose are taken for device calibration which includes prediabetic, diabetic and healthy samples. The baseline characteristics of the samples are shown in Table \ref{dataset1}.

\begin{table}[htbp]
	\caption{Baseline characteristics of collected samples for calibration, validation and testing.}
	\label{dataset1}
	\centering
	\begin{tabular}{p{3cm}p{2cm}p{2cm}p{2cm}p{2cm}}
		\hline
		Samples Basic &Capillary &Serum&Capillary &Serum\\
		Characteristics & Glucose & Glucose& Glucose & Glucose\\
		\hline
		& \multicolumn{2}{c}{\textbf{Calibration}}&\multicolumn{2}{c}{\textbf{Validation and Testing}}\\
		\hline
		\hline
		Age (Years)& \multicolumn{4}{c}{\textbf{Prediabetic Samples}}\\
		Male:-18-80& Male:-23&Male:-13&Male:-18 & Male:-10 \\
		Female:-17-75&Female:-20&Female:-16&Female:-16 &Female:-09\\
		\hline
		Age (Years)&\multicolumn{4}{c}{\textbf{Diabetic Samples}}\\
		Male:-18-80&Male:-30&Male:-18&Male:-14&Male:-15\\
		Female:-17-75&Female:-19&Female:-12&Female:-12 &Female:-12\\
		\hline
		Age (Years) &\multicolumn{4}{c}{\textbf{Healthy Samples}}\\
		Male:-18-80&Male:-09&Male:-08&Male:-07& Male:-05\\
		Female:-17-75&Female:-12&Female:-07&Female:-07 &Female:-08\\
		\hline
		Age (Years)&\multicolumn{4}{c}{\textbf{Total Samples}}\\
		Male:-18-80&Male:-62&Male:-39&Male:-39& Male:-30\\
		Female:-17-75&Female:-51&Female:-35&Female:-35& Female:-29\\
		\hline
	\end{tabular}
\end{table}

\begin{figure}[htbp]
	\centering
	\includegraphics[width=0.85\textwidth]{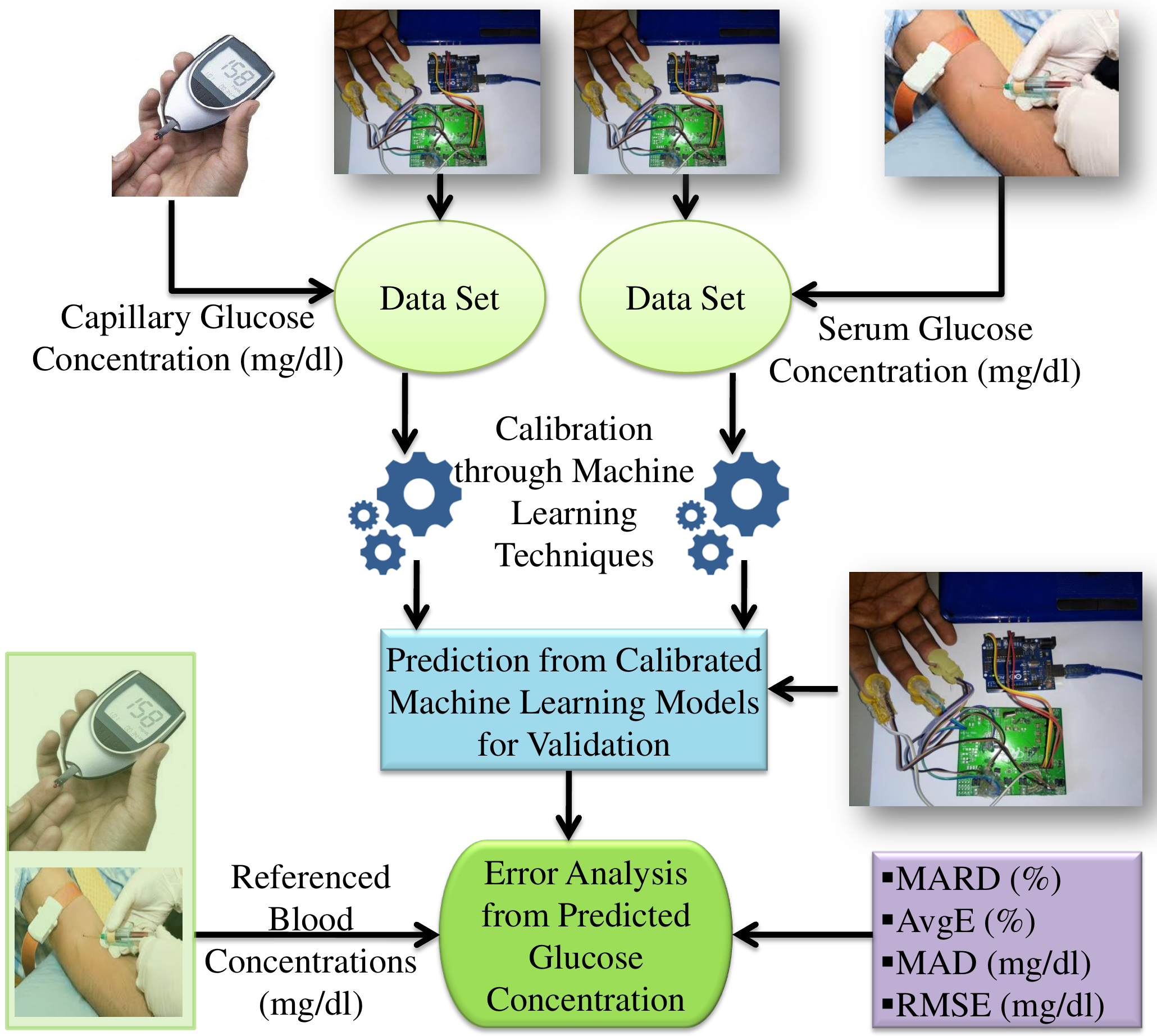}
	\caption{The process flow of calibration and validation of proposed device (iGLU 2.0).}
	\label{ML_flow}
\end{figure}

The proposed process flow of calibration and validation is shown in Figure \ref{ML_flow}. Mean absolute deviation (MAD), AvgE, mARD, and Root Mean Square Error (RMSE) are calculated to analyze the performance of proposed device. 

\subsection{Proposed Deep Neural Network (DNN) and Other Models for Glucose Sensor Calibration}

\begin{figure}[htbp]
	\centering
	\includegraphics[width=0.8\textwidth]{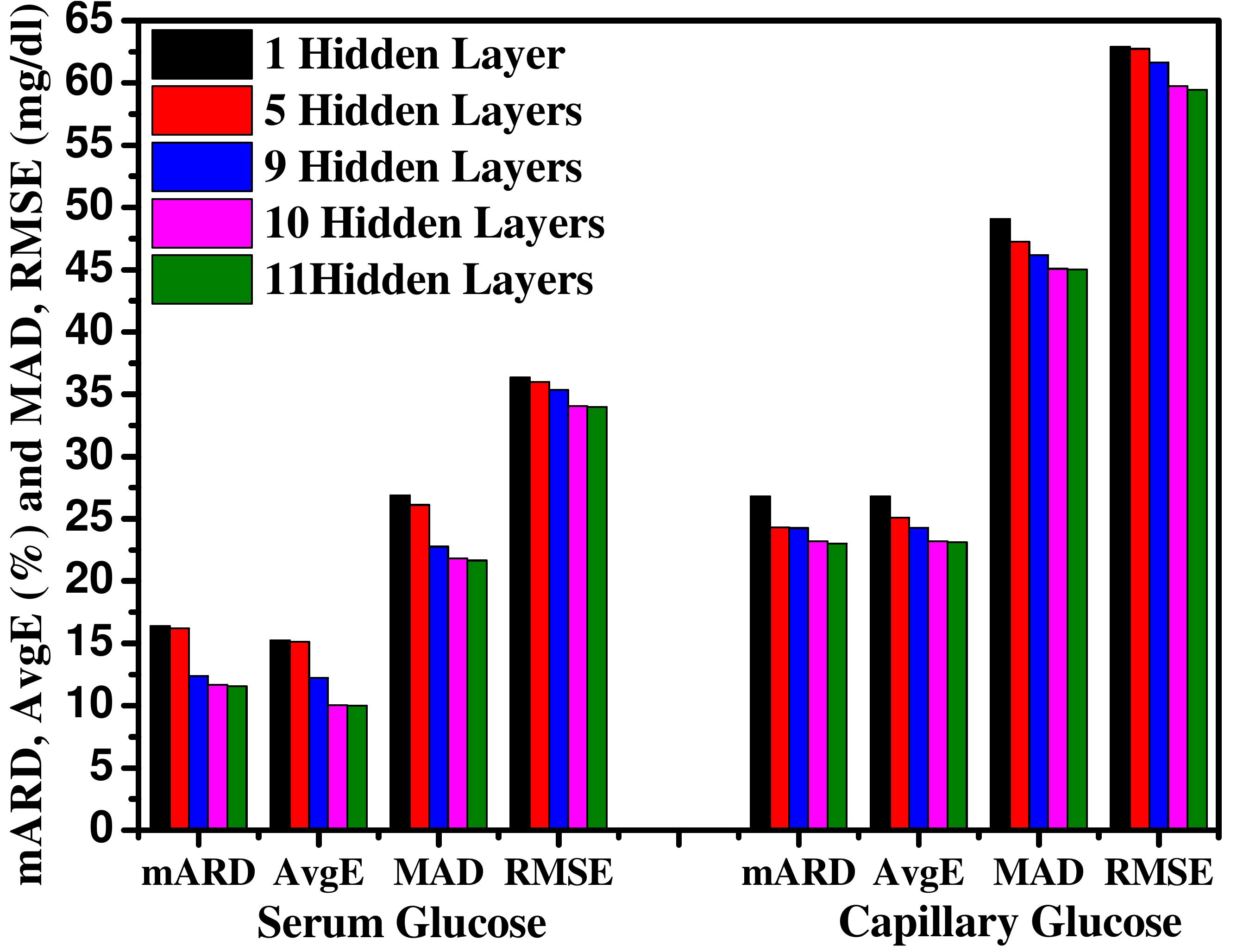}
	\caption{Error analysis of DNN models using different hidden layers}
	\label{HL_analysis}
\end{figure}

\begin{figure}[htbp]
	\centering
	\includegraphics[width=0.85\textwidth]{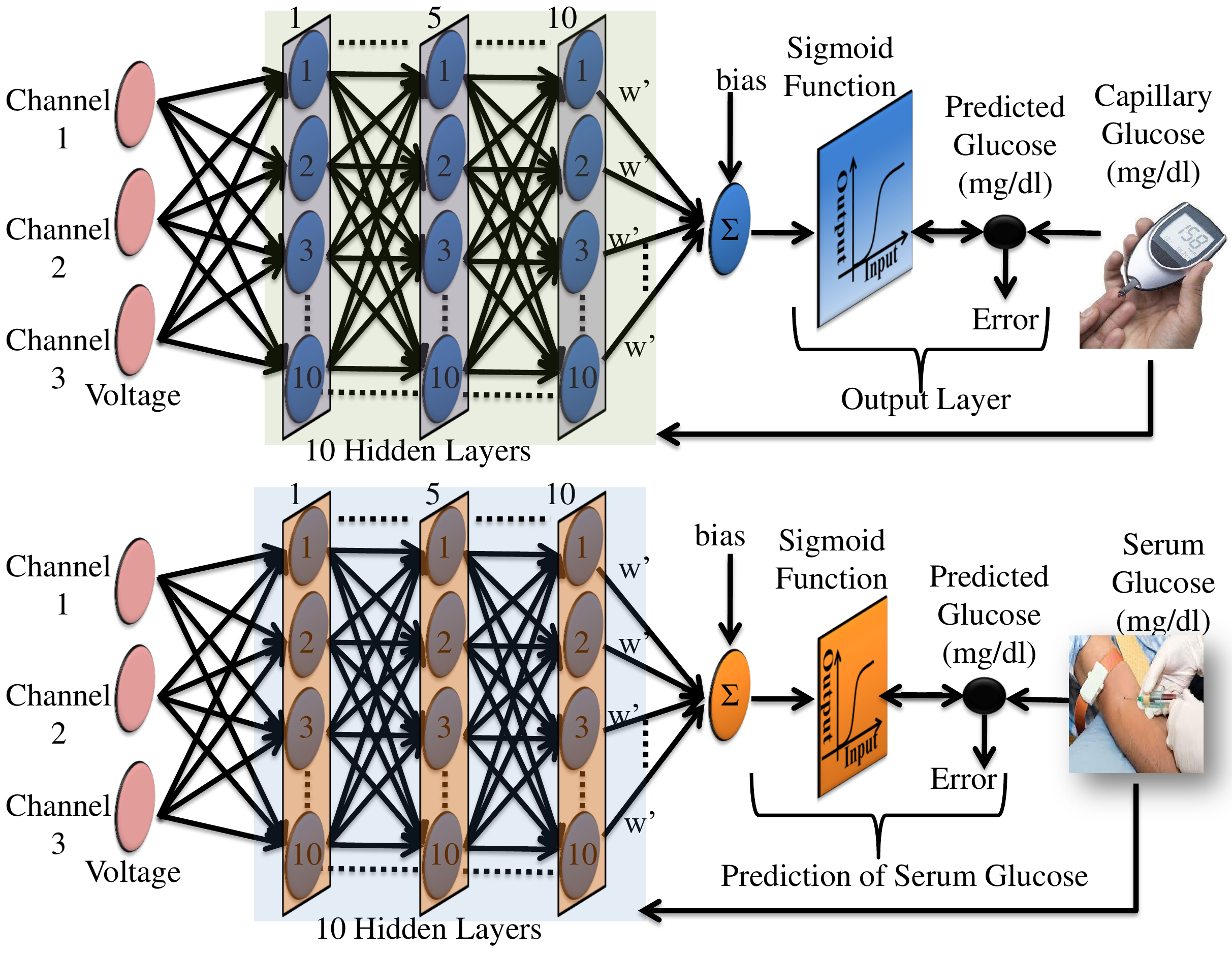}
	\caption{The Deep Neural Network (DNN) for proposed work}
	\label{dnnfit}
\end{figure}

Several machine learning-based computation models have been examined to get optimized regression method in terms of precise measurement. Support vector regression with multiple Gaussian kernels is used to analyze the optimized model. In case of blood glucose estimation, a margin of tolerance (epsilon) is set in approximation as per the independent variables. To reduce the error, a kernel (medium gaussian kernel) scale value is also set to the square root of number of predictors. This customized kernel scale is used to remove the outlier from data. The cubic kernel (kernel with polynomial degree 3) based model is customized to remove the outliers. This customized kernel scale is used for precise measurement. Kernel (fine gaussian kernel) scale value is also set to the square root of predictors which is divided by 4 ($\frac{\sqrt{P}}{4}$). The fitting model of DNN has also been analyzed for capillary and serum glucose prediction in the current work \cite{Song2015}. Sigmoid activation functions have been used in the proposed DNN models and have been trained through Levenberg-Marquardt backpropagation algorithm \cite{Quesada_2020_5_algorithms_to_train_a_neural_network}. 
In proposed models, 10 hidden layers have been analyzed to predict precise glucose values. The error analysis is graphically represented in Figure \ref{HL_analysis}.
These optimized models have been used to predict the glucose values from logged data. Here, the responses (voltages) from three channels are taken as inputs of the proposed DNN models. The predicted blood glucose values corresponding to capillary and serum glucose are formed through the modelling of three channels voltage values. Weights of the voltage values correlate predicted glucose values to the channels data. Overall accuracy through DNN is better using 10 hidden layers. The block diagram of DNN model is given in Figure \ref{dnnfit}. 

\subsection{Proposed Method: Multiple Polynomial Regression (MPR) model of Glucose Concentration with Voltage}

\begin{figure}[htbp]
	\centering
	\includegraphics[width=0.85\textwidth]{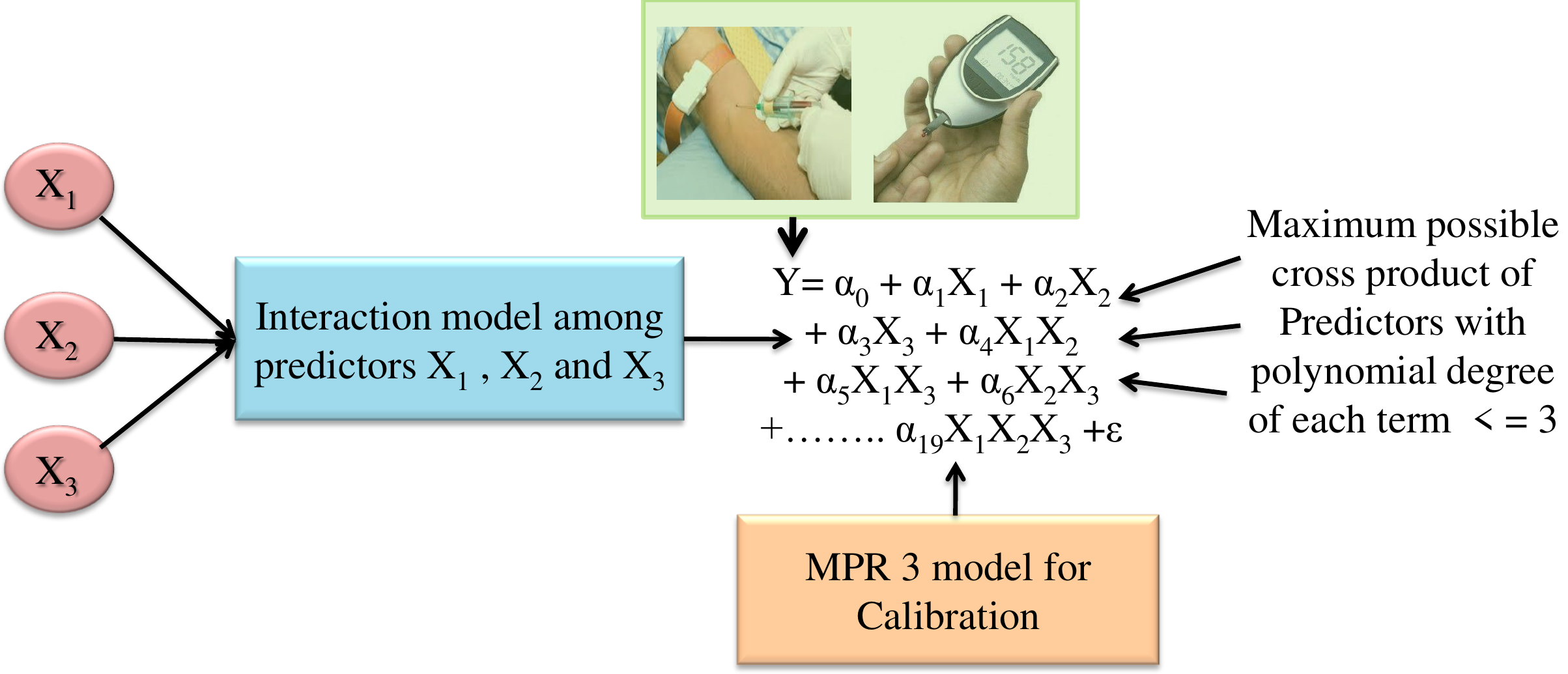}
	\caption{MPR 3 model representation for calibration.}
	\label{MPR3model}
\end{figure}

Multiple polynomial regression with polynomial degree 3 based model (MPR 3) is applied to analyze the optimized model for capillary and serum glucose prediction. Using multiple polynomial regression, proposed model represents the relationship between three channels detector's outputs and corresponding referenced glucose values (serum glucose or capillary glucose) by fitting a cubic kernel based regression model which is represented as follows:
\begin{eqnarray}
\nonumber
y & = & a_{1}x_{1}^{3}+a_{2}x_{2}^{3}+a_{3}x_{3}^{3}+a_{4}x_{1}^{2}x_{2}+a_{5}x_{1}^{2}x_{3}+a_{6}x_{1}x_{2}^{2} 
 +a_{7}x_{1}x_{3}^{2}+a_{8}x_{2}^{2}x_{3}\\
\nonumber
&& +a_{9}x_{2}x_{3}^{2}+a_{10}x_{1}^{2}+a_{11}x_{2}^{2}+a_{12}x_{3}^{2}
+a_{13}x_{1}x_{2}x_{3}+a_{14}x_{1}x_{2}+a_{15}x_{1}x_{3}\\
&&+a_{16}x_{2}x_{3}+a_{17}x_{1}+a_{18}x_{2}+a_{19}x_{3}+\epsilon,
\label{Eqn:Model}
\end{eqnarray}
where, $a_{1}$-$a_{19}$ are regression coefficients and $\epsilon$ is residual of errors. The model is used to observe the predicted glucose value from iGLU. The values of these regression coefficients and residual of error depend upon the predictors and corresponding response of calibrated model. In the proposed model, three channels voltage values are represented as $x_{1}$, $x_{2}$ and $x_{3}$ as independent variables (predictors) and estimated glucose (mg/dl) is the dependent variable $y$. Each value of detector's output voltage value is associated with reference glucose value.  Proposed multiple polynomial regression model (MPR 3) kernel for calibration is represented in Figure \ref{MPR3model}.

\begin{figure*}[htbp]
	\centering
	\subfigure[LSVR for capillary glucose]{\includegraphics[width=0.4\textwidth]{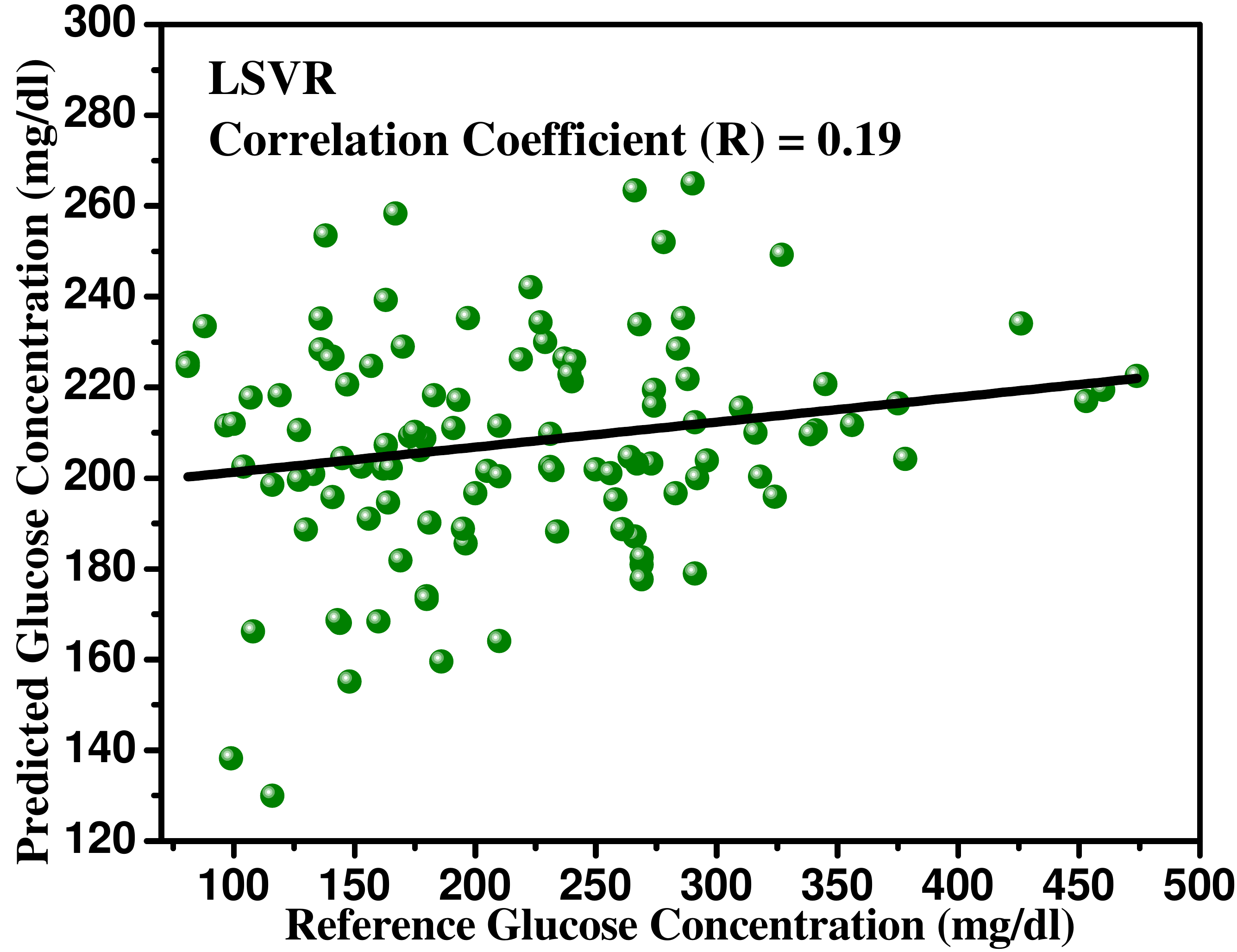}}\label{}
	\subfigure[FGSVR for capillary glucose] {\includegraphics[width=0.4\textwidth]{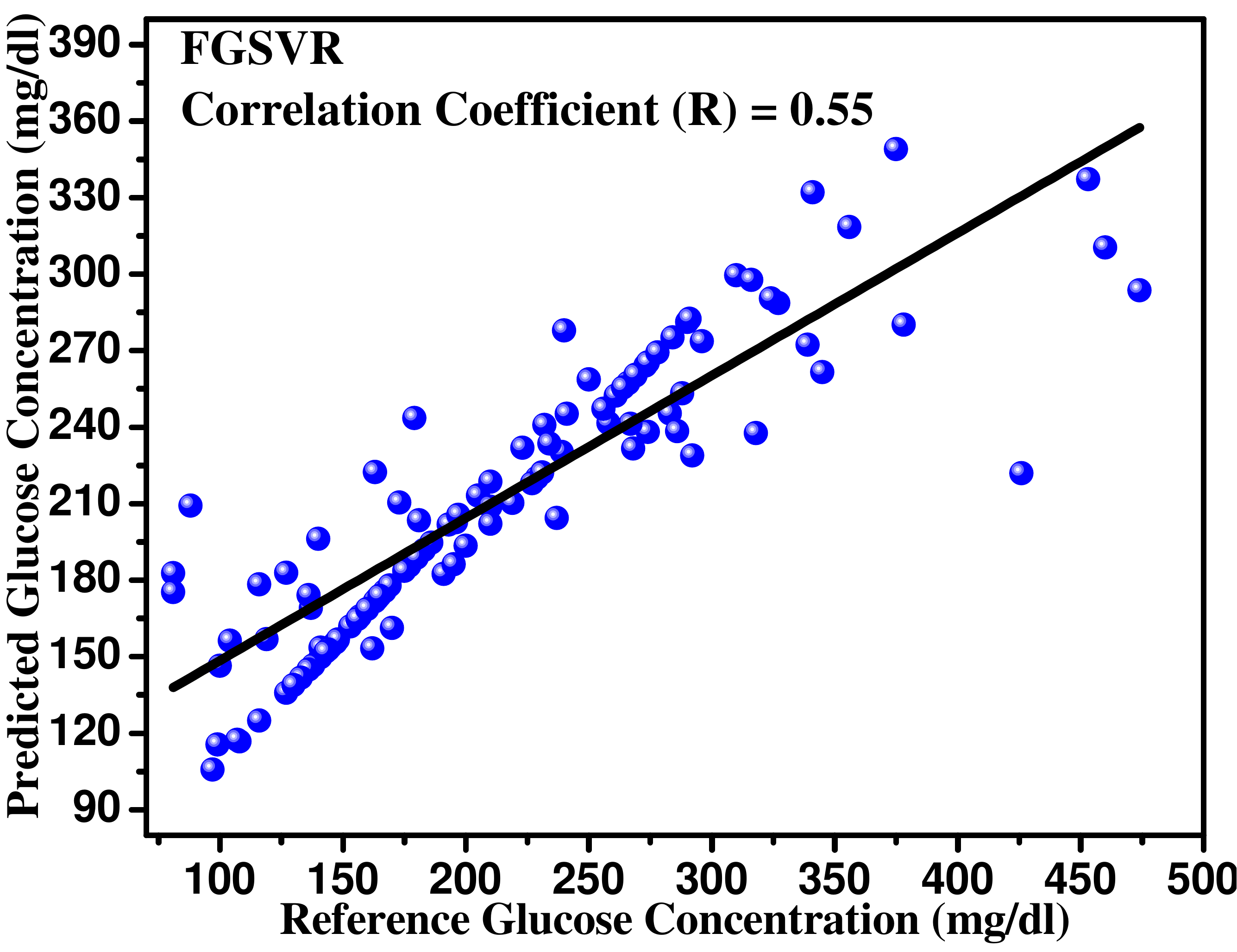}}\label{}
	\subfigure[DNN for capillary glucose]
	{\includegraphics[width=0.39\textwidth]{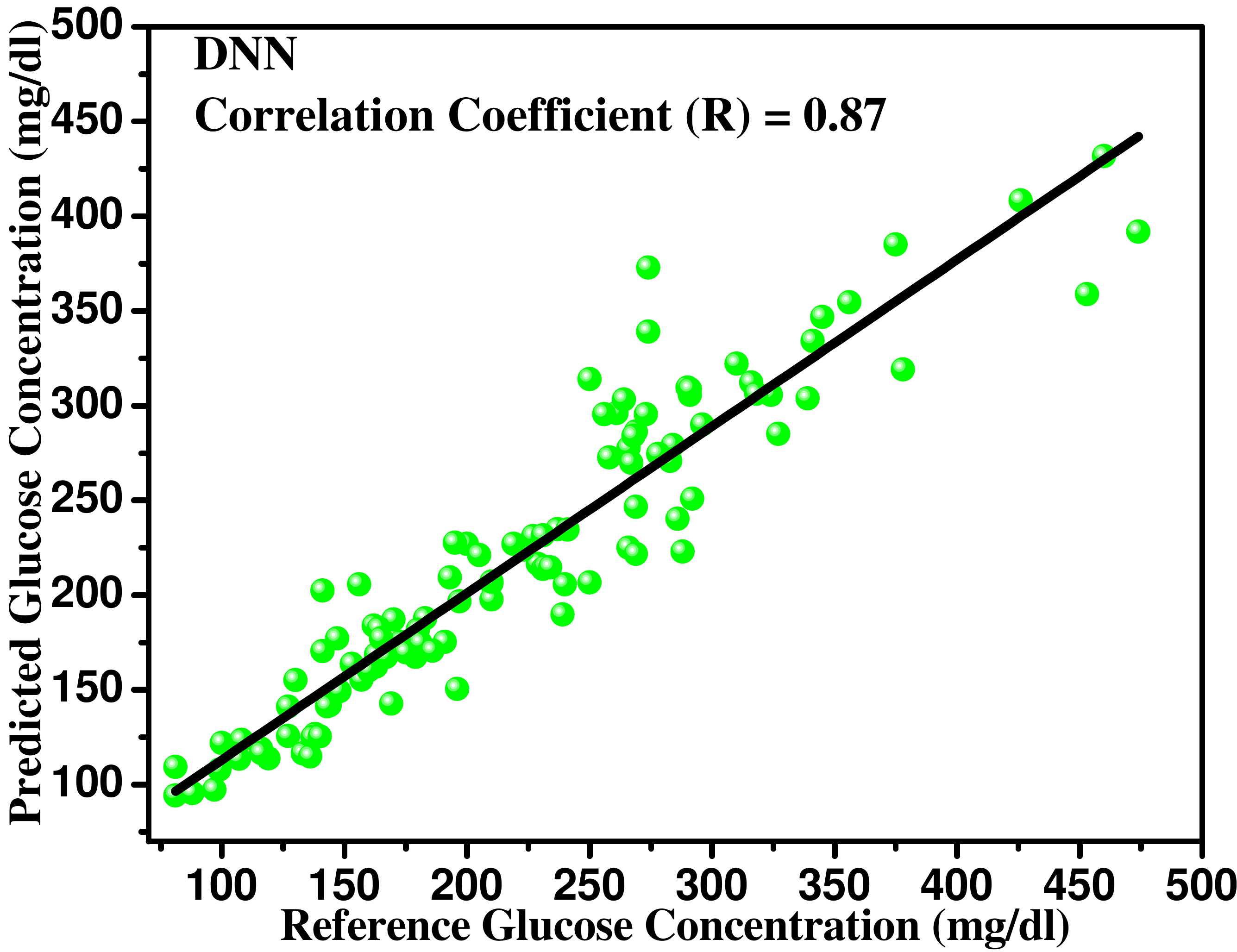}}\label{}
	\subfigure[MPR3 for capillary glucose]
	{\includegraphics[width=0.39\textwidth]{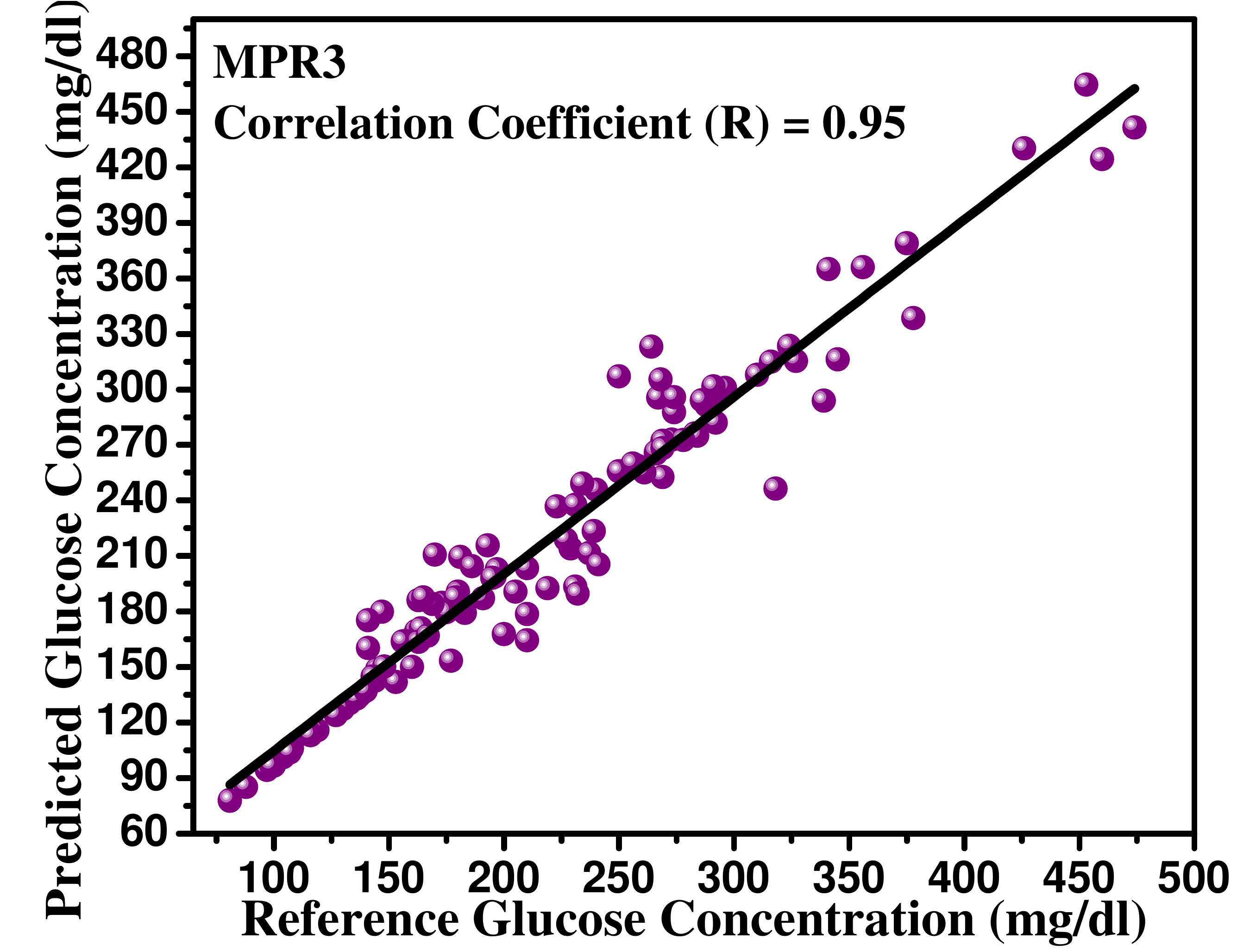}}\label{}
	\subfigure[LSVR for serum glucose]
	{\includegraphics[width=0.39\textwidth]{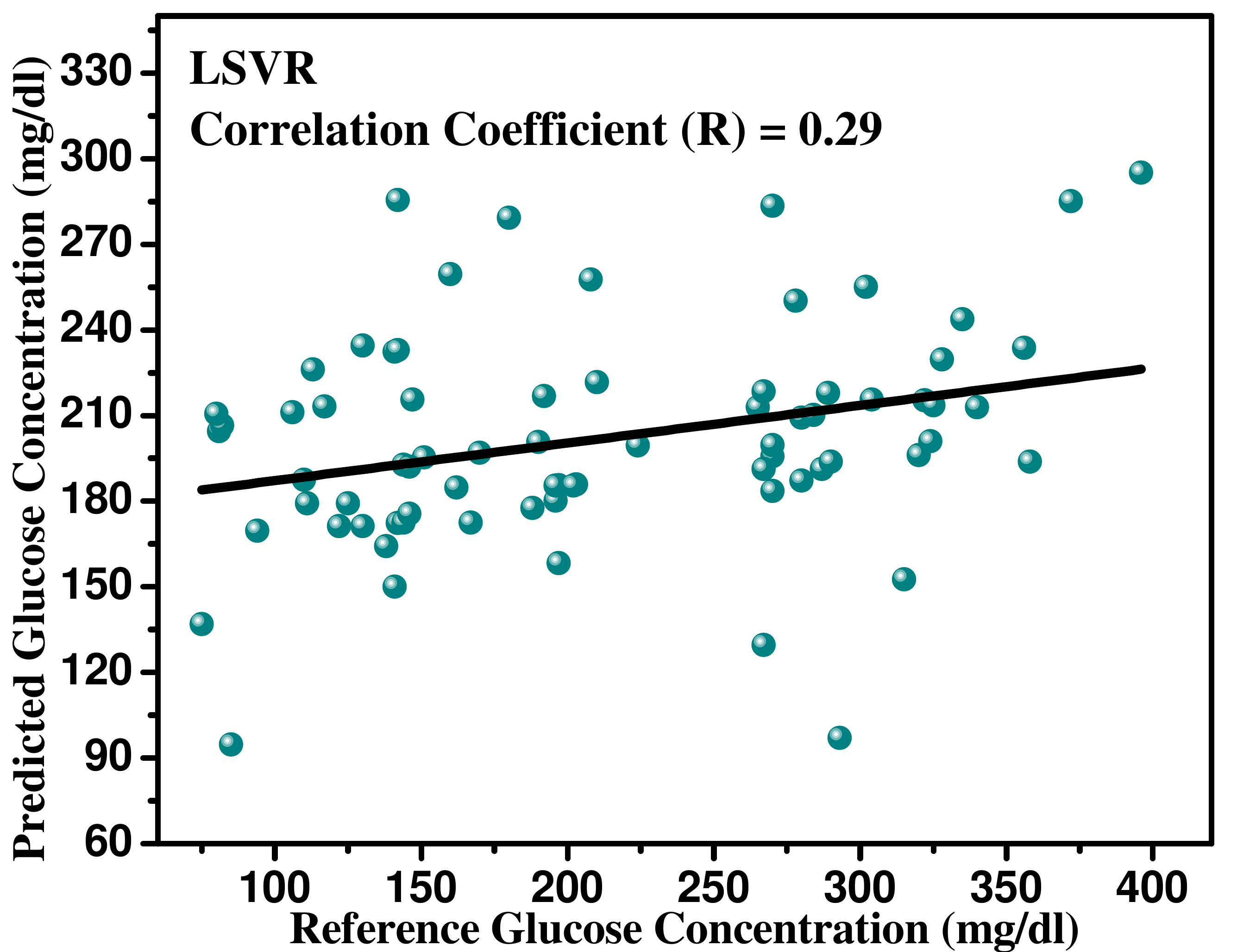}}\label{}
	\subfigure[FGSVR for serum glucose]
	{\includegraphics[width=0.39\textwidth]{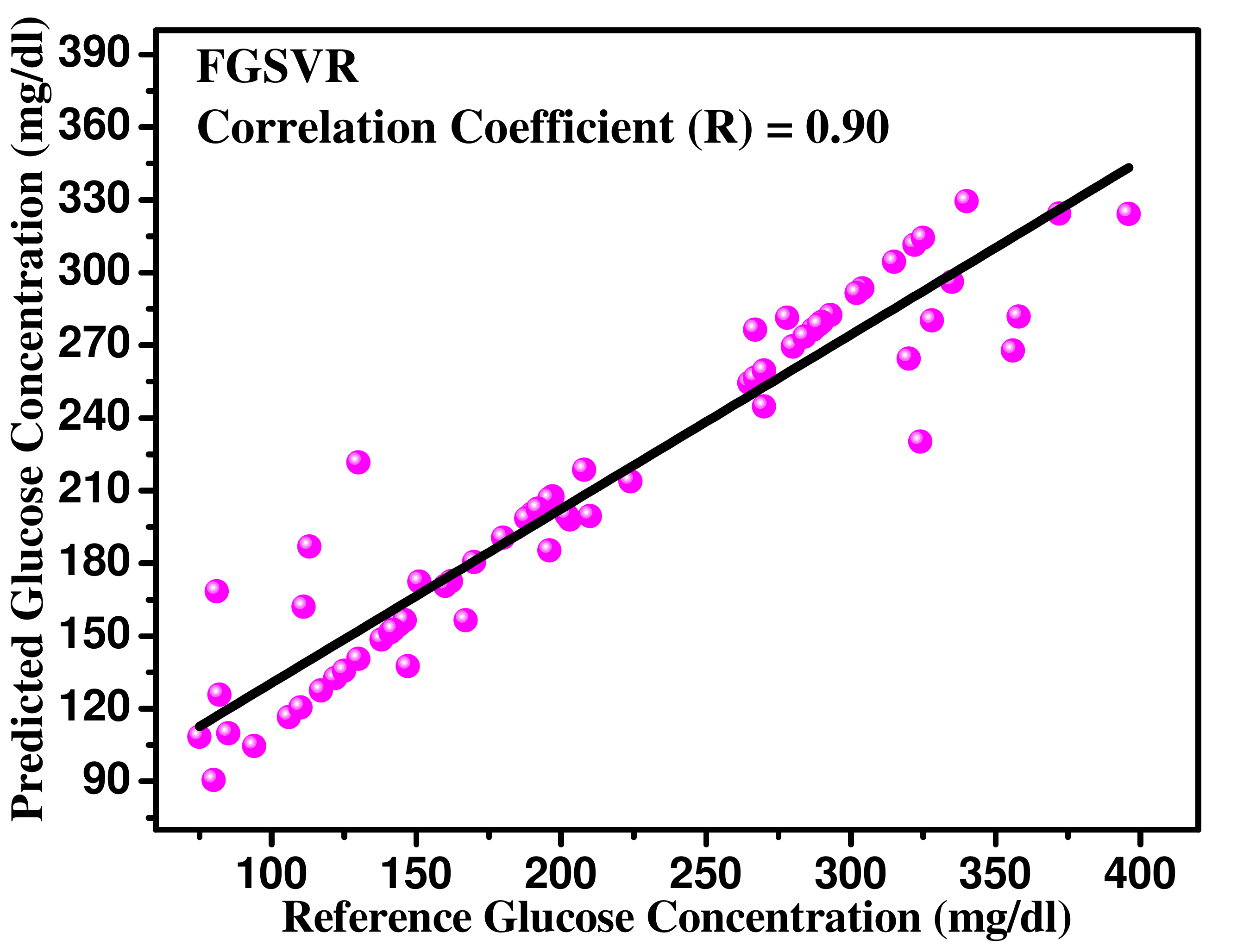}}\label{}
	\subfigure[DNN for serum glucose]
	{\includegraphics[width=0.39\textwidth]{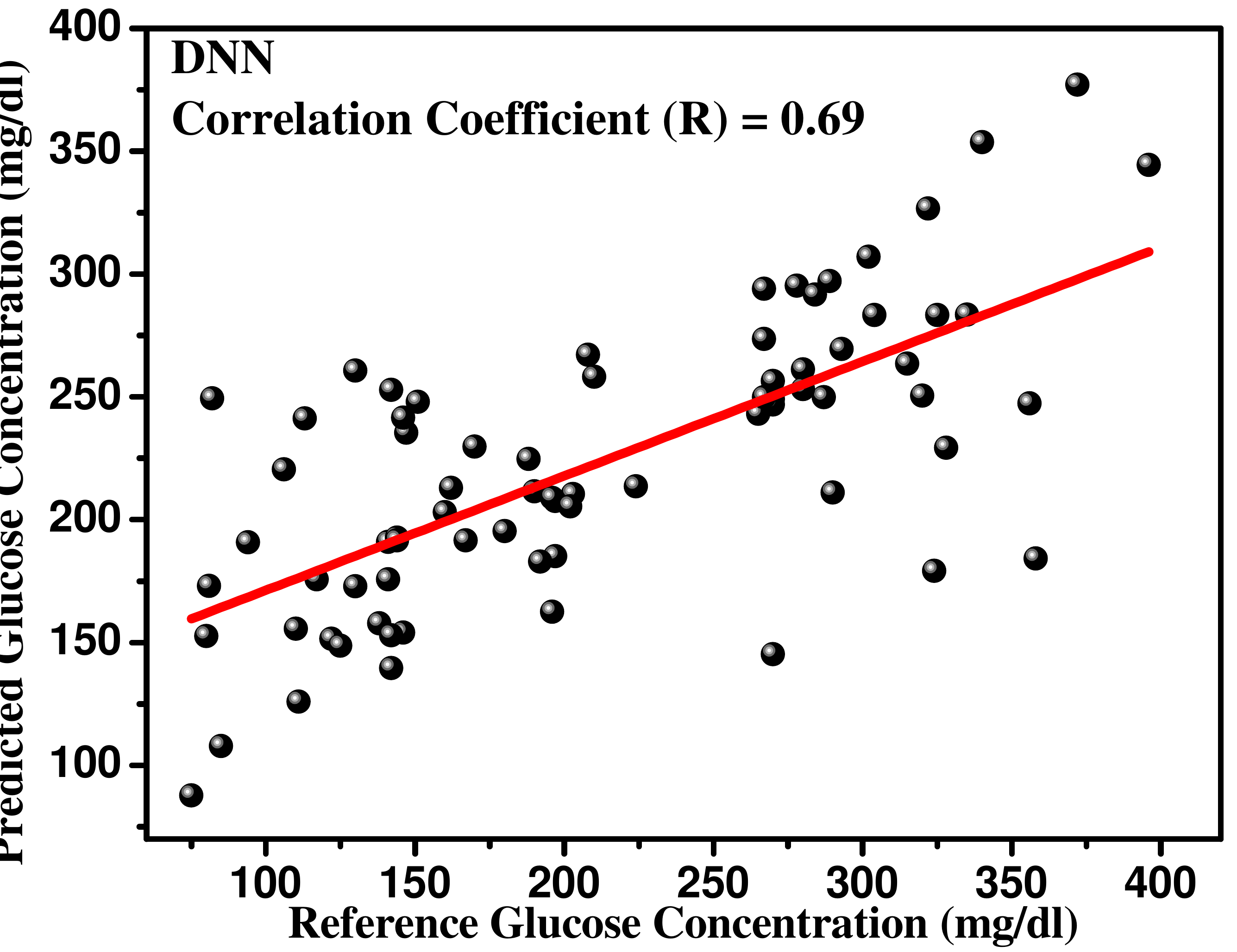}}\label{}
	\subfigure[MPR3 for serum glucose]
	{\includegraphics[width=0.39\textwidth]{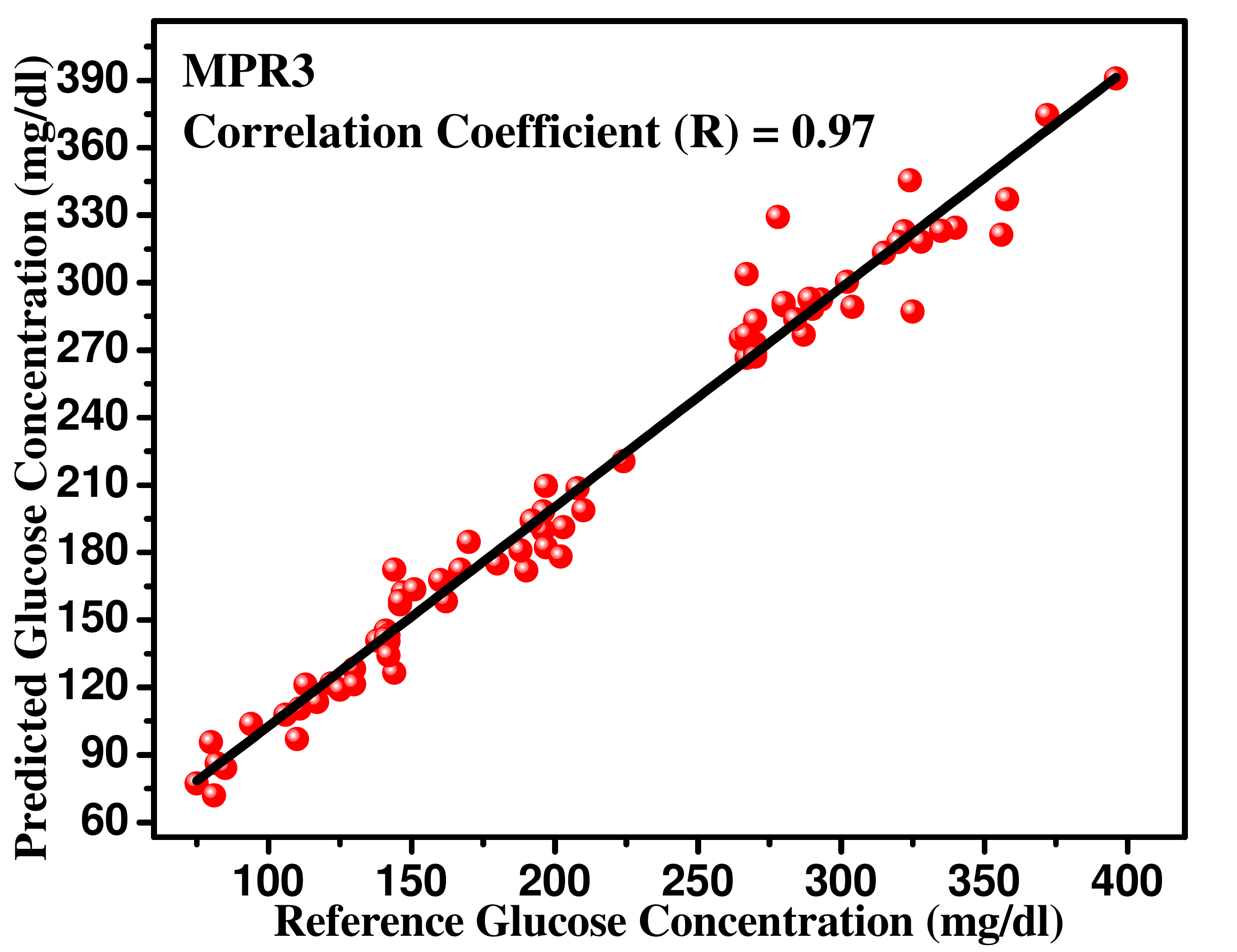}}\label{}
\caption{Correlation plot of predicted and referenced blood glucose concentration model during calibration.} 
	\label{wave}
\end{figure*}

Proposed MPR is a complex multivariable interaction approach based regression model. 19 customized interacted variables combinations based kernel is an optimized model which is represented in eq. \ref{MPR3model}. The maximum polynomial degree 3 of terms (interactions of multiple variables) have been analyzed for precise model. The correlation plots between predicted glucose concentration and reference glucose concentration for regression models are represented in Figure \ref{wave}(a)- Figure \ref{wave}(h). The error analysis of calibrated machine learning regression models is also represented in Table \ref{TBL:calibration}.
 
\begin{table}[htbp]
	\caption{Statistical Analysis of calibration of proposed model and existing techniques.}
	\label{TBL:calibration}
	\centering
	\begin{tabular}{ccccc}
		\hline
		Regression&mARD&AvgE&MAD&RMSE\\
		Model&\%&\%&mg/dl&mg/dl\\
		\hline
		\hline
				Linear SVR (Capillary) &34.40&31.27&65.64&83.50\\
				Serum &39.24&36.21&70.59&83.21\\
				\hline
		Cubic SVR (Capillary) &31.85&27.32&59.42&79.66\\
		Serum &26.69&32.55&51.92&73.32\\
		\hline
			Quadratic SVR (Capillary) &33.43&29.73&63.59&81.38\\
				Serum &33.42&10.86&61.35&83.47\\
				\hline
		Medium Gaussian SVR (Capillary) &31.36&26.82&58.43&77.83\\
		Serum &26.50&24.66&47.75&66.01\\
		\hline
			Coarse Gaussian SVR (Capillary) &33.71&30.74&64.58&82.10\\
				Serum &40.09&34.75&70.37&81.05\\
				\hline
		Fine Gaussian SVR (Capillary) &14.31&12.49&27.36&45.06\\
		Serum &12.31&10.45&20.96&31.09\\
		\hline
		DNN (Capillary)&29.06&22.14&46.47&62.51\\
		Serum &9.11&8.95&19.47&27.95\\
		\hline
		MPR3 (Capillary) &6.07&6.09&13.28&19.71\\
		Serum &4.86&4.88&9.42&13.57\\
		\hline
	\end{tabular}
\end{table}

Proposed MPR 3 model represents better results of calibration and validation compared to DNN based model and other regression models because of its complexity and polynomial interaction-based approach with polynomial degree 3. During analysis, prediction of serum glucose is found more precise compared to prediction of capillary glucose using MPR 3.

\section{Design of the Proposed Novel Glucometer for Serum Glucose Monitoring - The iGLU 2.0}
\label{Sec:Proposed-Work}

The proposed device is based on short wave NIR spectroscopy using wavelengths (940 and 1300 nm) and implemented using three channels. Each channel contains emitter and detector of a particular wavelength for glucose detection. The flow of proposed work is represented in Figure \ref{FIG:Proposed_Glucometer_Overview}.

\subsection{The Proposed Approach for Data Acquisition}

\begin{figure}[htbp]
	\centering
	\includegraphics[width=0.85\textwidth]{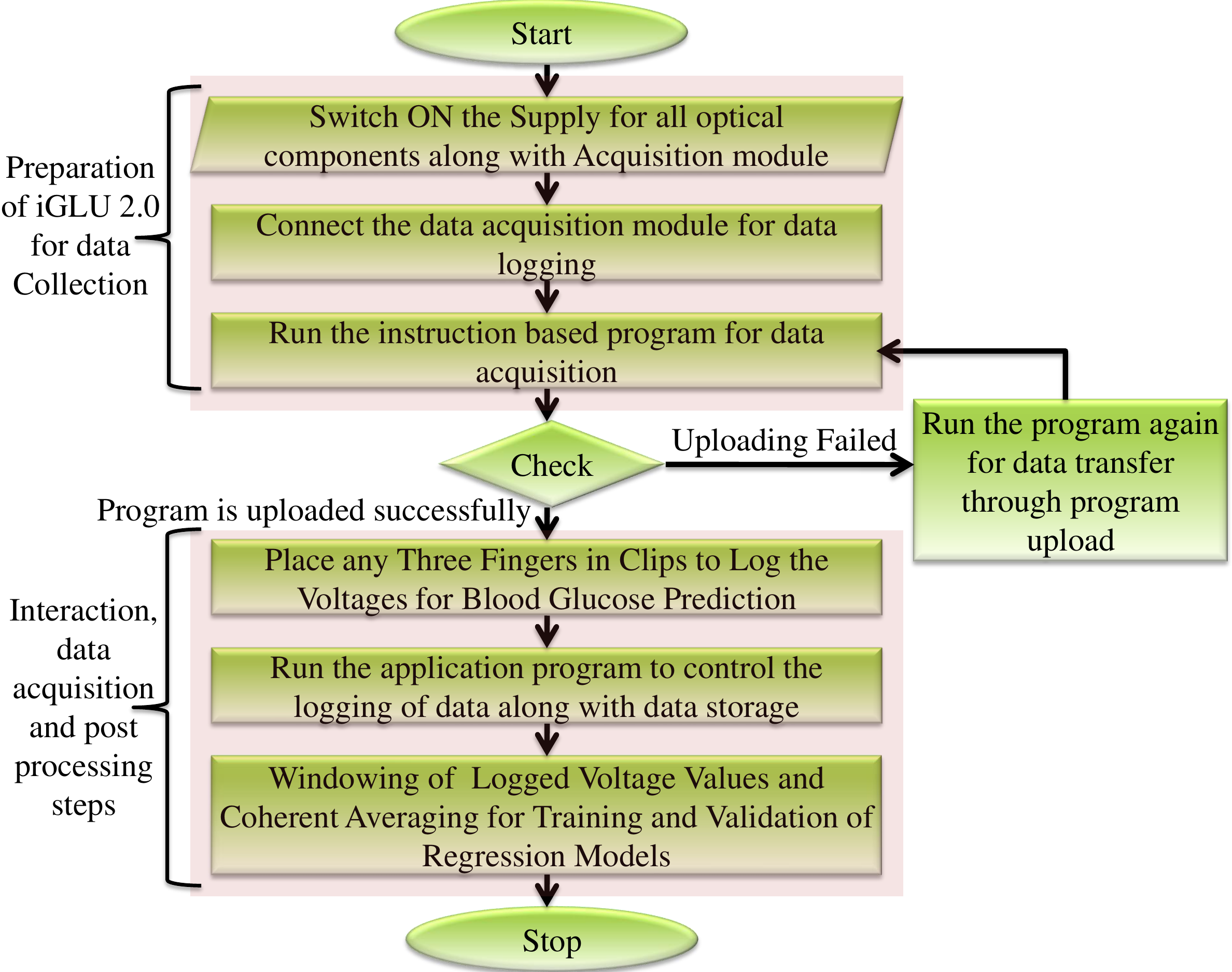}
	\caption{Process flow data acquisition for proposed device (iGLU 2.0).}
	\label{pseudocode}
\end{figure}

The pseudocode of data acquisition for proposed iGLU 2.0 is represented in process flow which is shown in Figure \ref{pseudocode}.
The data is collected and serially logged by 16 bit ADC with sample rate of 128 samples/sec. The logged data is calibrated and validated through an optimized model of existing regression techniques for precise measurement. 

Independent samples of person aged 24-50 have also been taken for testing and validation of iGLU 2.0. The serum glucose values have been stored on the cloud using open IoT platform. The data can be accessed by patients and doctor. On the basis of serum glucose values, the treatment can be provided at a remote location. A 2 layer PCB is designed and system is developed to embed NIR LEDs and detectors.


The circuit is designed to implement the setup for optical detection \cite{Jain_IEEE-MCE_2020-Jan_iGLU1}. Emitters, detectors and ADS 1115 are embedded with optimized biasing with 5 V DC supply. Passive components have been chosen for better efficiency of NIR LEDs and detectors. All NIR LEDs and detectors are connected with proper biasing. The detectors are used in photoconductive mode. All detectors have specifications of daylight blocking filters \cite{Jain_IEEE-MCE_2020-Jan_iGLU1}. ADS 1115 has been used to serially transferred the three channels data in decimal form. The ADS 1115 is controlled through microcontroller ATmega328P using arduino uno board.

\subsection{The Proposed System of the Glucometer}

\begin{figure}[htbp]
	\centering
	\includegraphics[width=0.85\textwidth]{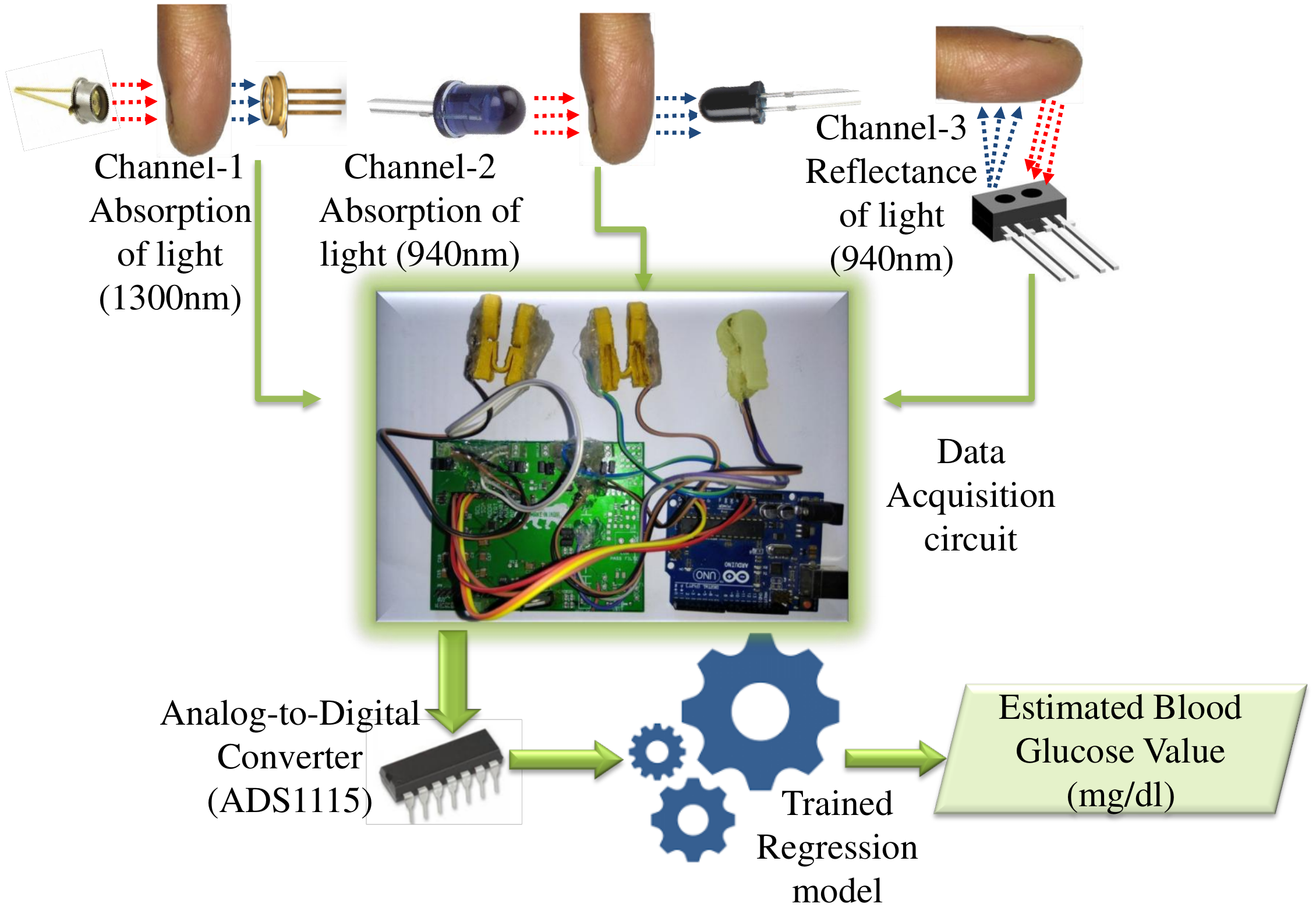}
	\caption{Block Representation of proposed device (iGLU 2.0).}
	\label{fig:Proposed_Glucose_Measurement_System}
\end{figure}

In this paper, we implement optical detections using specific wavelengths for absorptions and reflection. Concentration of glucose molecules depends upon the change in light intensity. The logged voltage values from the detectors depend on received light intensity. During placing the object (fingertip or earlobe) between NIR LED and detector, the voltages (data) are logged. The coherent averaging of voltage samples is done to improve the measurement error. For calibration and validation, the averaging of 1024 samples has been done in 8 seconds. Block representation of proposed iGLU is shown in Figure \ref{fig:Proposed_Glucose_Measurement_System}. During experimental analysis, serum glucose has been measured in laboratory and capillary glucose has been measured by invasive glucometer SD check gold device for validation \cite{seo2009clinical}. The serum and capillary glucose values are taken as referenced glucose values (mg/dl). At that time, detector responses (in milivolts) have been collected through all channels simultaneously. During measurement process, the data has been logged through all channels in the form of voltages from detector of each channel using ADS 1115. These logged voltage values are corresponding to serum and capillary glucose values. 

%
%
%

\section{Validation of the Proposed Device iGLU 2.0}
\label{Sec:Experimental-Results}

\subsection{Testing and Error Analysis}

After calibration of iGLU 2.0, 50 different healthy, prediabetic and diabetic samples of capillary glucose and 37 samples of serum glucose have been taken to validate the device. During device validation, 26 males and 20 females samples are tested following medical protocols. The samples are taken in fasting, post-prandial and random modes for validation and testing. The baseline characteristics and error analysis is represented in Table \ref{dataset1} and \ref{TBL:validation}.

\begin{table}[htbp]
	\caption{Statistical Analysis of validation of proposed model and existing techniques.}
	\label{TBL:validation}
	\centering
	\begin{tabular}{ccccc}
		\hline
		Regression&mARD&AvgE&MAD&RMSE\\
		Model&\%&\%&mg/dl&mg/dl\\
		\hline
		\hline
			FGSVR (Capillary) &14.09&11.45&25.20&41.18\\
				Serum &9.17&9.12&19.09&27.34\\
				\hline
		DNN (Capillary) &23.19&22.14&45.07&59.74\\
		Serum &11.67&10.02&21.81&34.05\\
		\hline
		MPR3 (Capillary) &7.74&7.70&16.08&22.46\\
		Serum &5.009&4.97&9.74&12.98\\
		\hline
	\end{tabular}
\end{table}

\begin{figure}[htbp]
	\centering
	\subfigure[Samples of capillary glucose]{\includegraphics[width=0.45\textwidth]{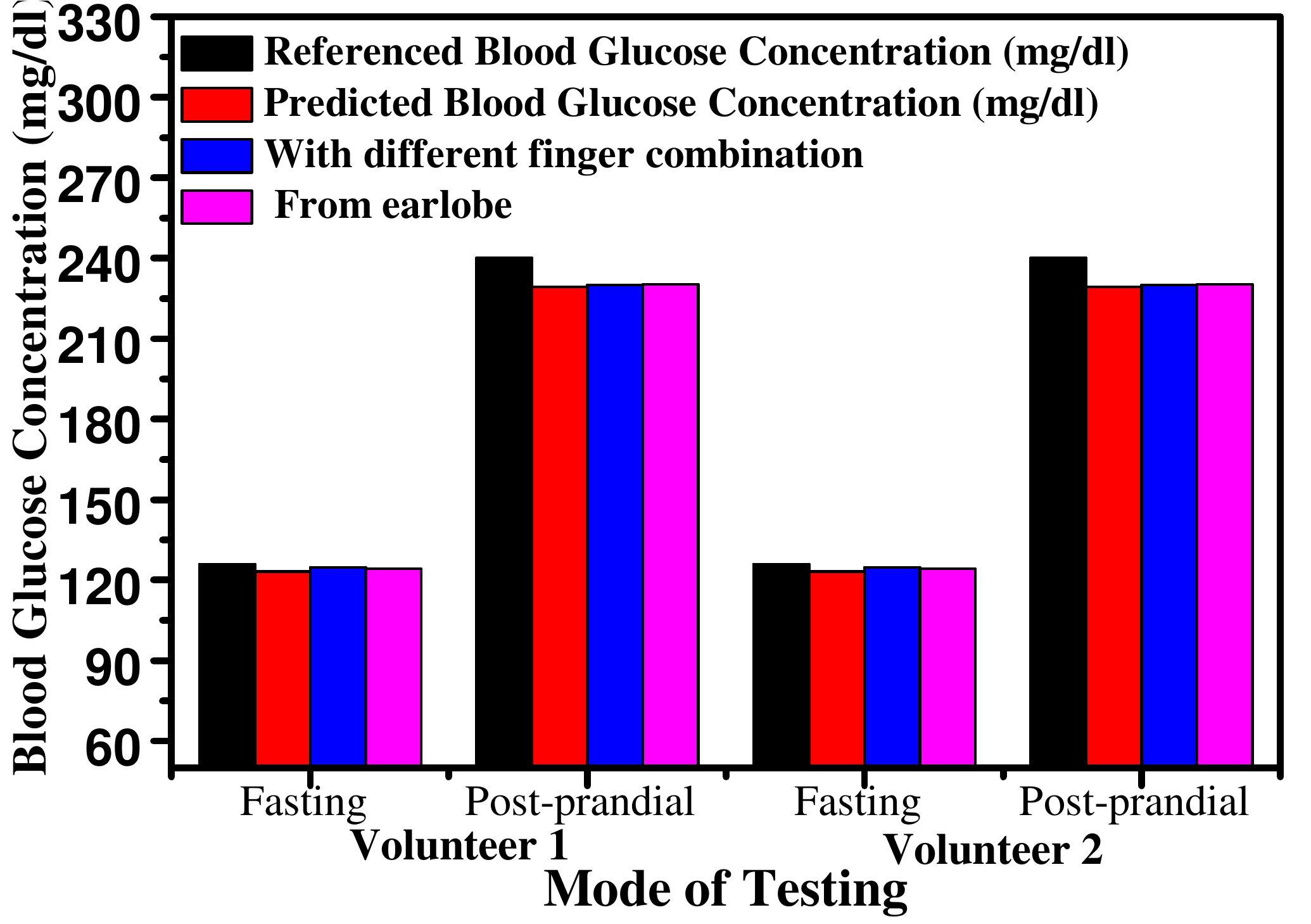}}\label{}
	\subfigure[Samples of serum glucose]{\includegraphics[width=0.45\textwidth]{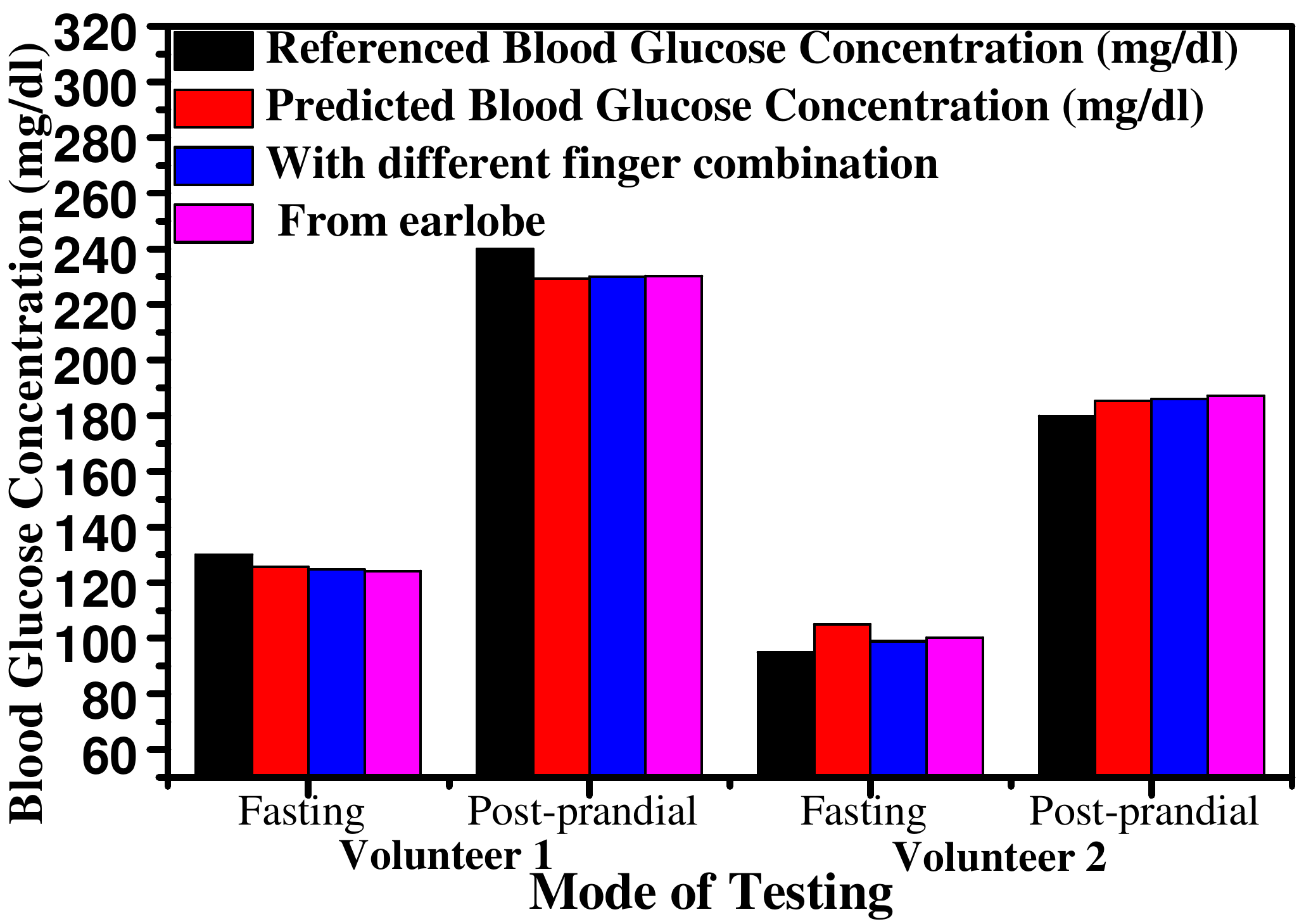}}\label{}
	\caption{Predicted and reference blood glucose concentration for validation of iGLU 2.0 on 2 volunteers.}
	\label{graph}
\end{figure}

As per analysis of results, serum glucose measurement is found more accurate compared to capillary glucose measurement using MPR 3 model. To test the device stability, an experimental analysis has been done with multiple measurements of 2 volunteers. For this experimental work, each volunteer has been chosen for measurement of capilary and serum glucose through iGLU 2.0. The measurement has been done in fasting and post-prandial mode. Result analysis is shown in Figure \ref{graph}.

\begin{figure}[!h]
	\centering
	\includegraphics[width=0.8\textwidth]{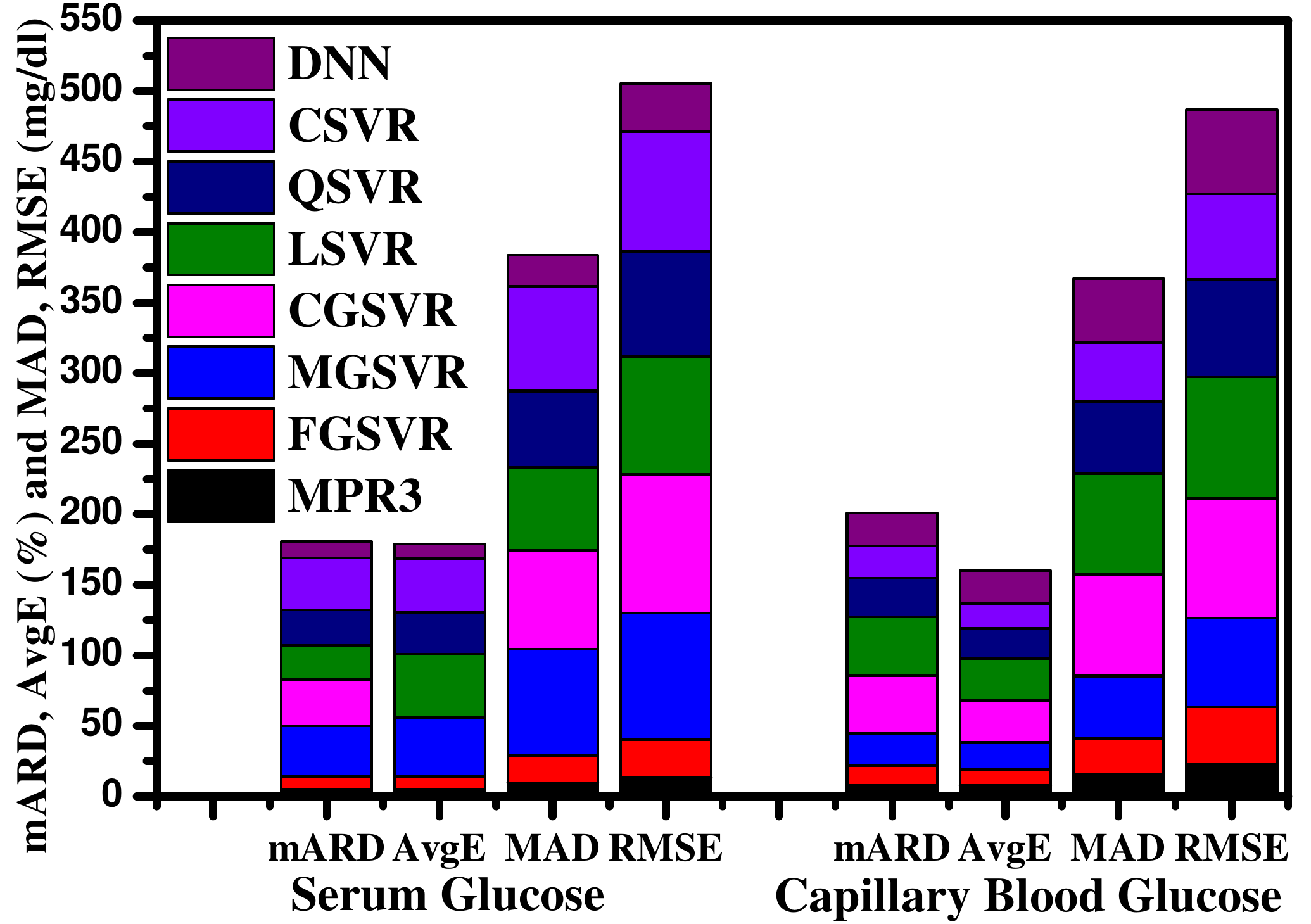}
	\caption{Error analysis of validation of data using existing regression techniques.}
	\label{validation}
\end{figure}

\subsection{Clarke Error Grid (CEG) Analysis}

\begin{figure}[!h]
	\centering
	\subfigure[Male samples of capillary glucose]{\includegraphics[width=0.4\textwidth]{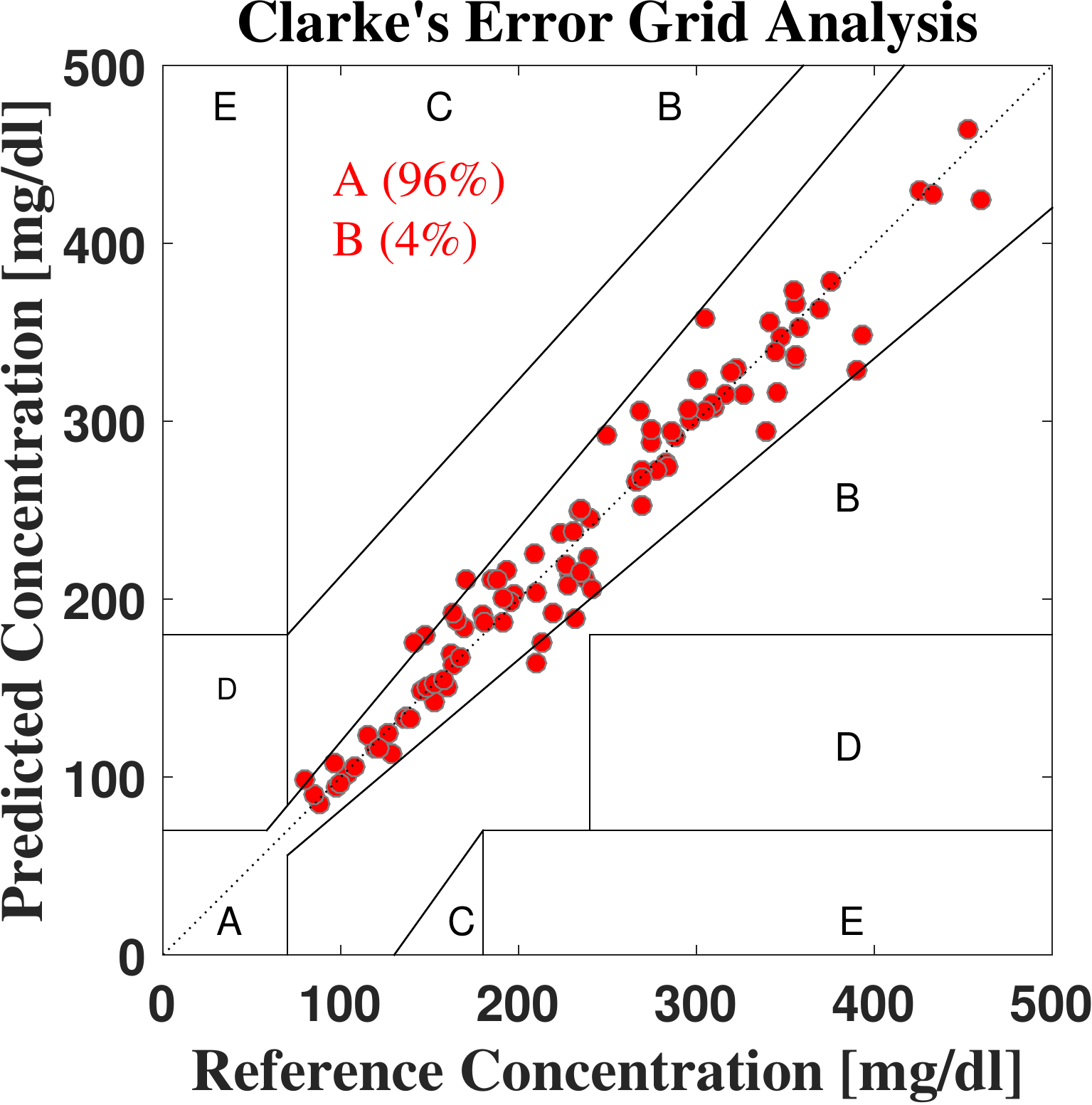}}\label{}
	\subfigure[Male samples of serum glucose]{\includegraphics[width=0.4\textwidth]{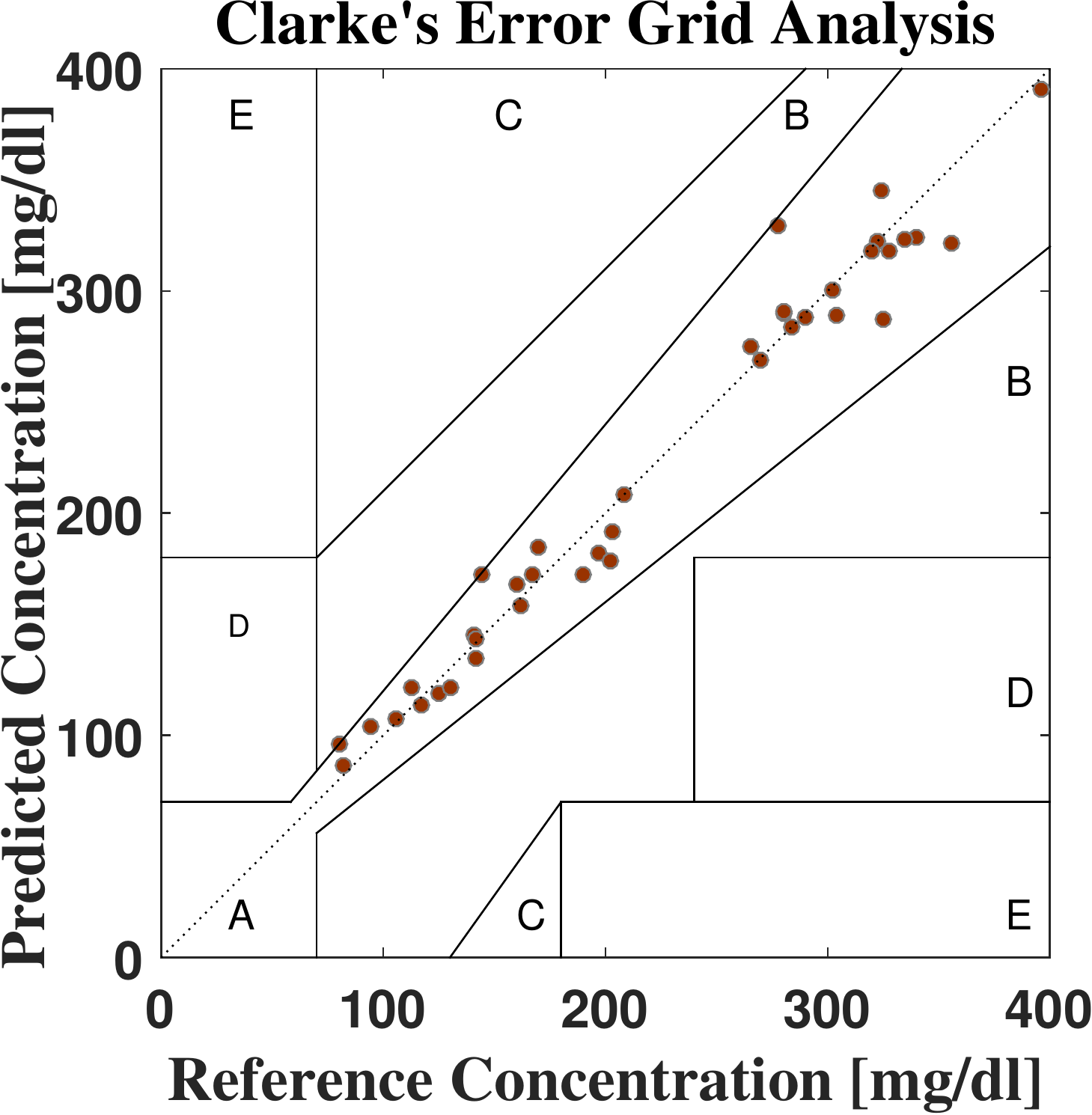}}\label{}
	\subfigure[Female samples of capillary glucose]{\includegraphics[width=0.4\textwidth]{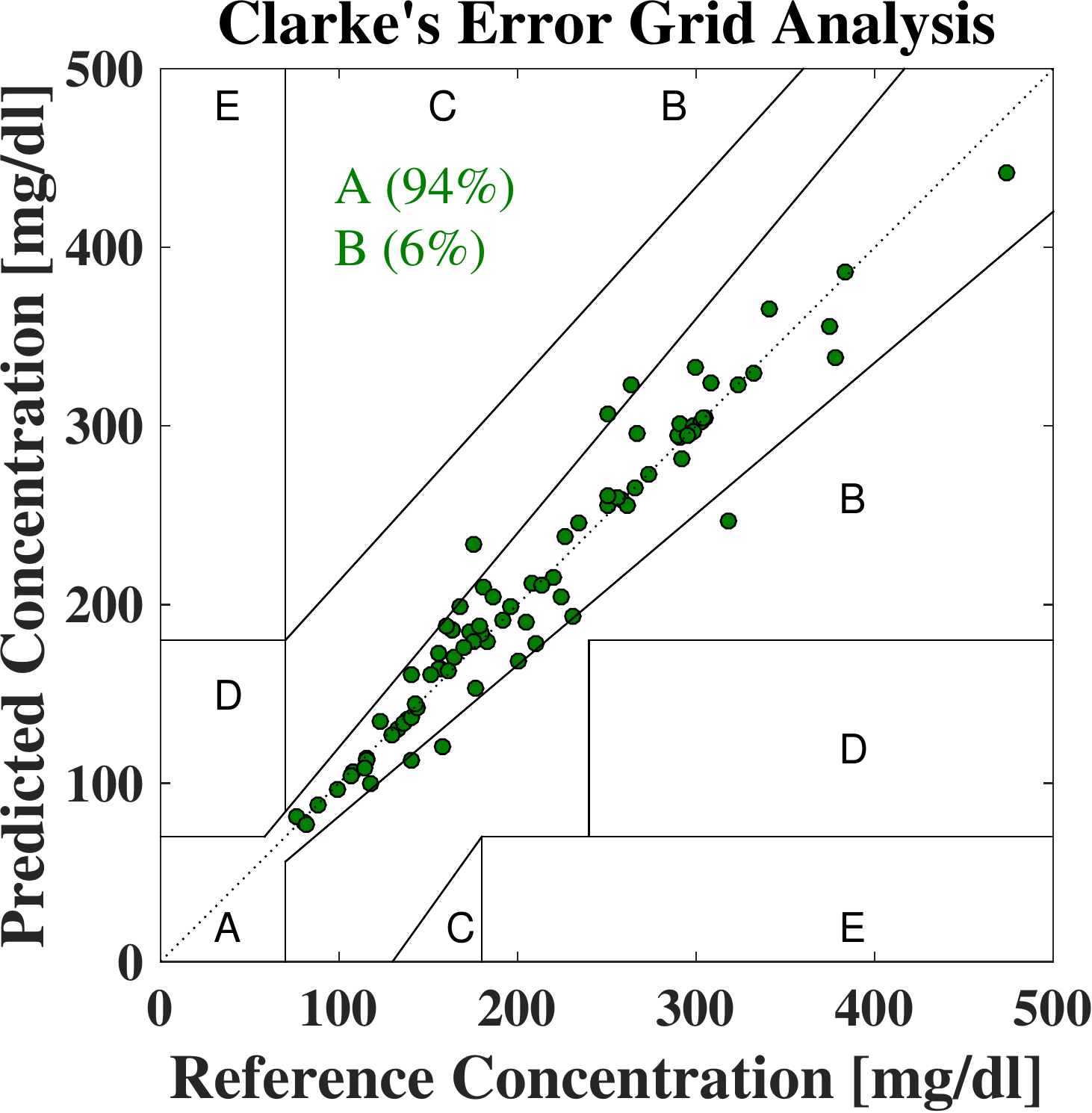}}\label{}
	\subfigure[Female samples of serum glucose samples]{\includegraphics[width=0.4\textwidth]{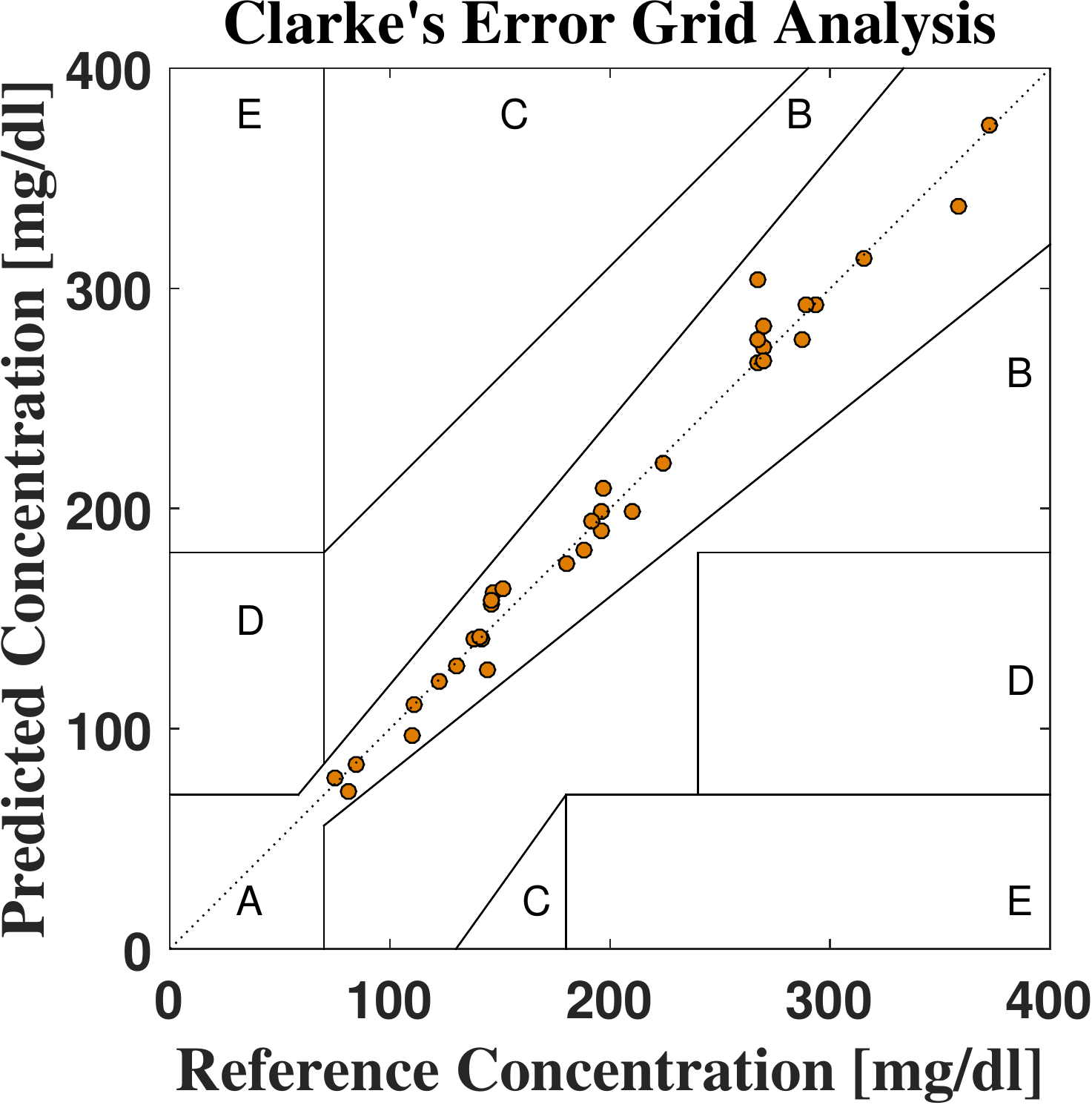}}\label{}
	\caption{CEG analysis of predicted blood glucose concentration for validation of iGLU 2.0 device.} 
	\label{gender}
\end{figure}

\begin{figure}[htbp]
	\centering
	\subfigure[Validation samples of capillary glucose]{\includegraphics[width=0.4\textwidth]{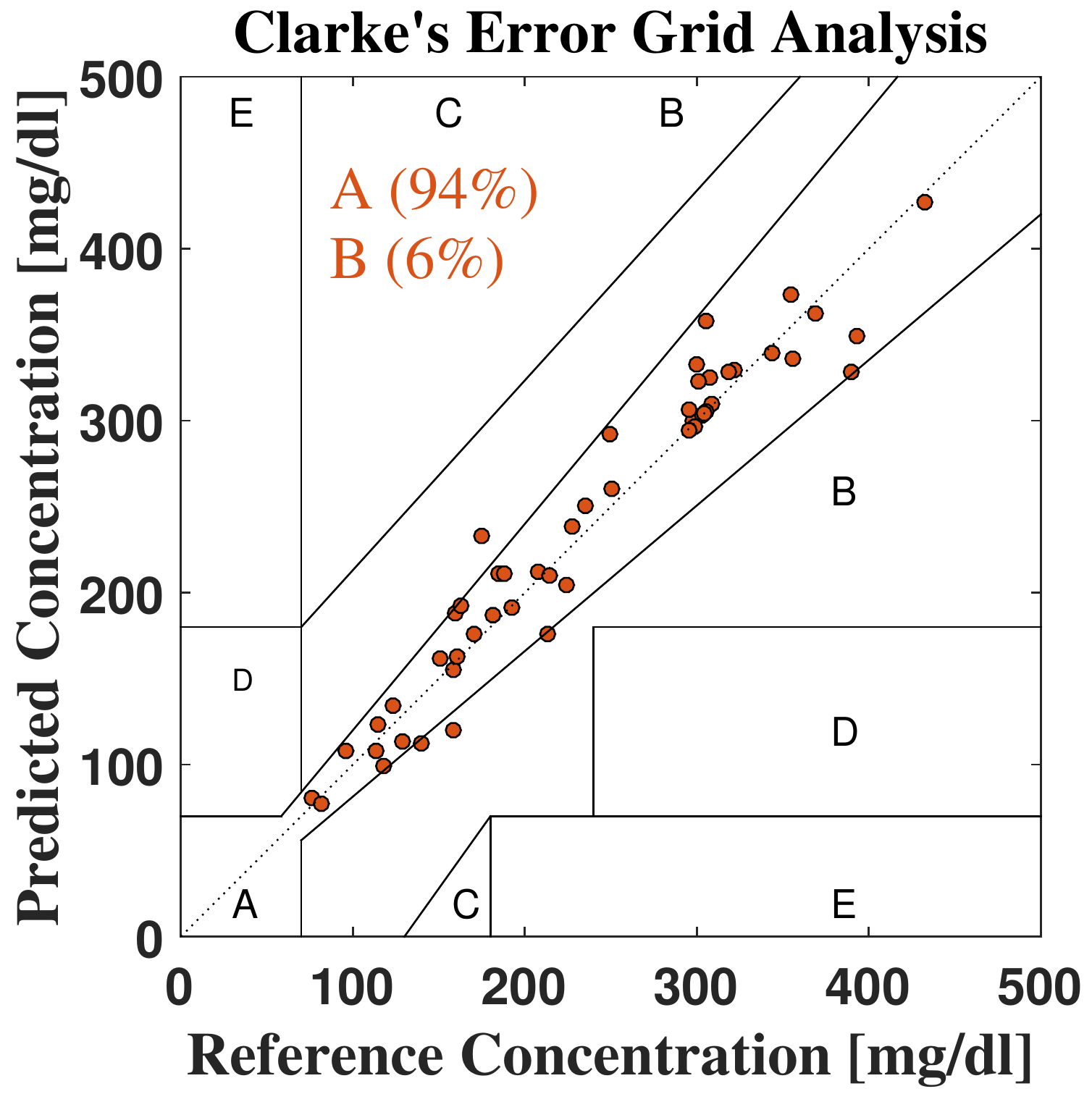}}\label{}
	\subfigure[Validation samples of serum glucose]{\includegraphics[width=0.4\textwidth]{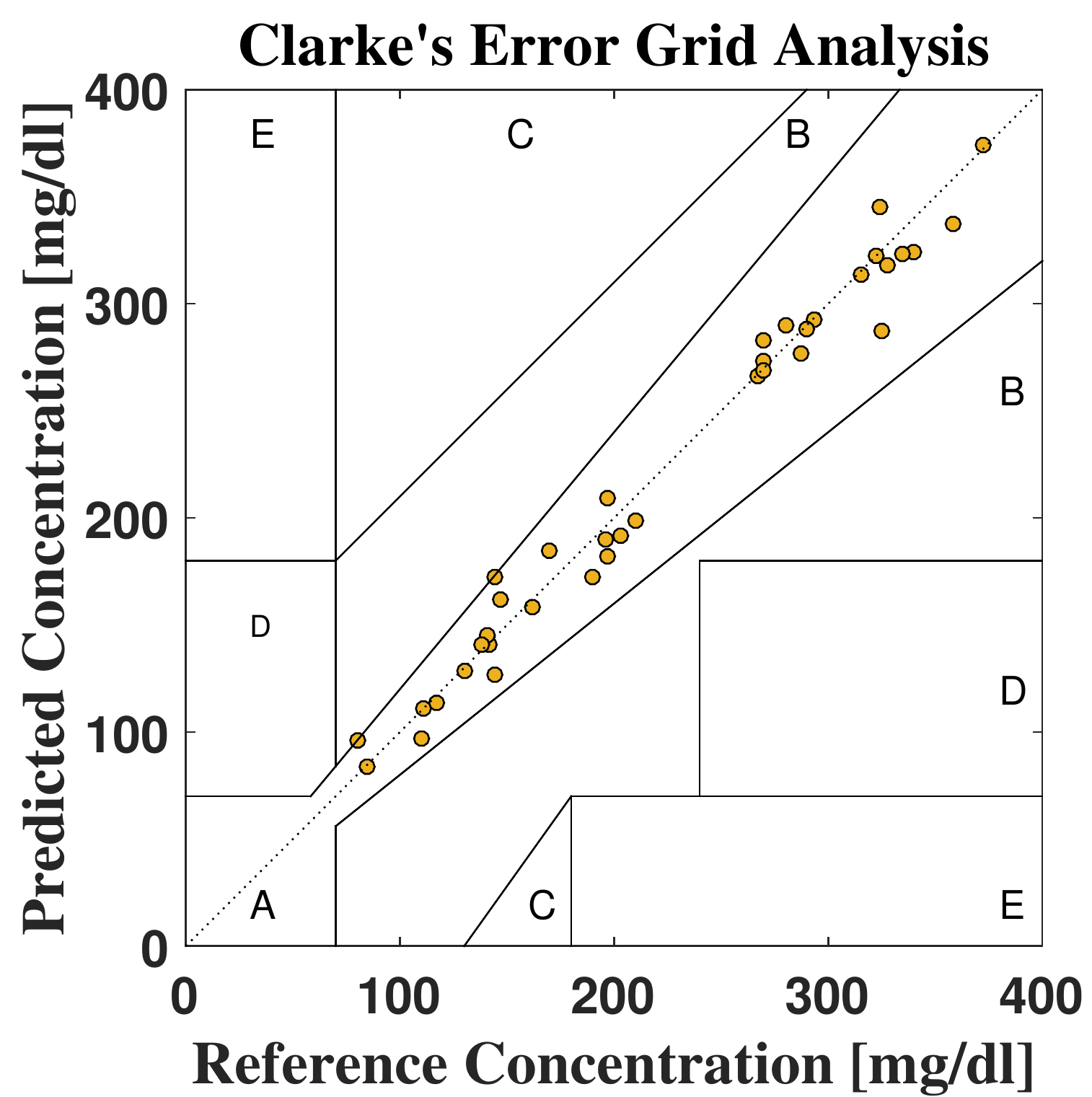}}\label{}
	\subfigure[Testing samples of capillary glucose]{\includegraphics[width=0.4\textwidth]{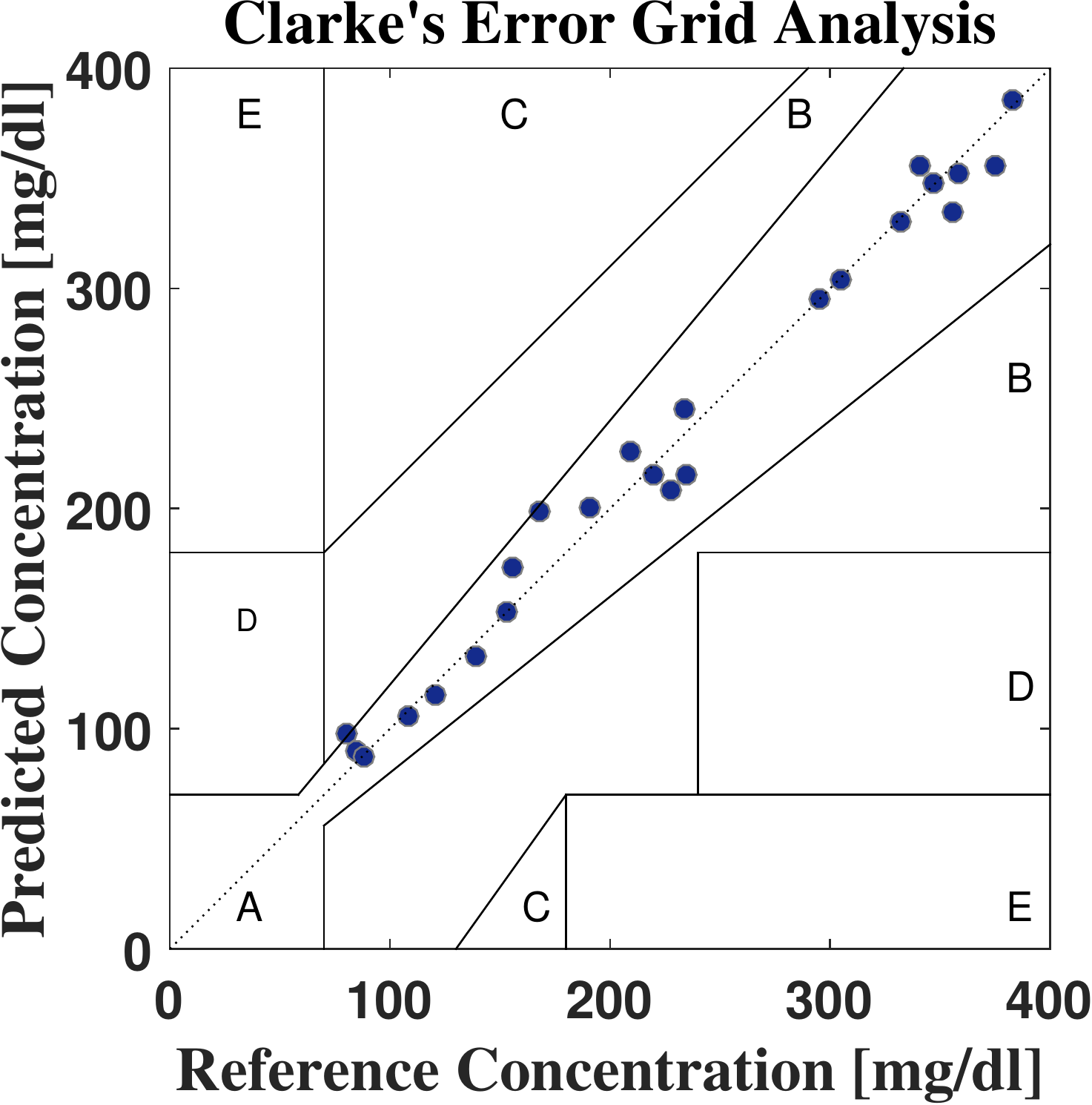}}\label{}
	\subfigure[Testing samples of serum glucose samples]{\includegraphics[width=0.4\textwidth]{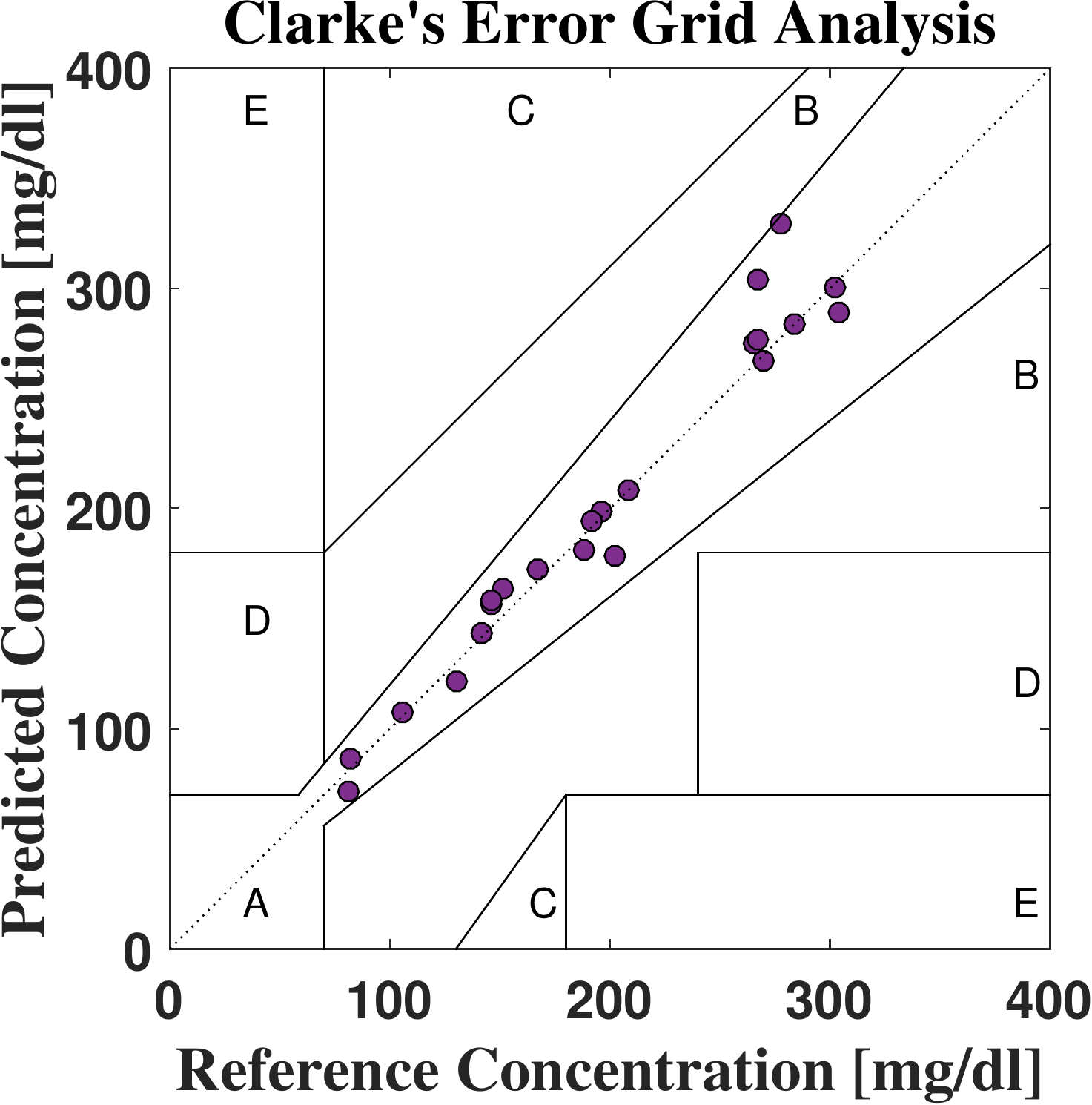}}\label{}
	\caption{CEG analysis of predicted blood glucose concentration for validation of iGLU 2.0 device.} 
	\label{cl_test}
\end{figure}

CEG analysis is explored by W. L. Clarke in 1970. This is used to analyze the accuracy of predicted blood glucose concentration values from proposed device. 
This error analysis has been presented to analyze the precision level of predicted blood glucose values from device iGLU 2.0 \cite{Clarke2005}. Clarke analysis explores the zones by the variation between referenced and predicted glucose concentration. All samples have been arranged gender-wise to confirm the accuracy of iGLU 2.0 for capillary and serum glucose measurement along with testing and validation which are represented in Figure \ref{gender} and \ref{cl_test} respectively. The proposed device is considered as desirable for clinical purpose. According to Table \ref{table_example}, proposed non-invasive device iGLU 2.0 is more precise compared to other glucose monitoring devices. 
During analysis, it has been observed that all samples of serum glucose exist in zone A and all samples of capillary glucose exist in zone A and zone B.

It has also been concluded that 4.86\% $mARD$ of serum glucose is better than 6.07\% $mARD$ of capillary glucose. The $AvgE$ and $MAD$ 4.88\% and 9.42 mg/dl represent the precision level of the proposed device. Real-time analysis and cross validation have been performed for real-time applications \cite{Jain_arXiv_2019-Nov30-1911-04471_iGLU1}.

\begin{table*}[htbp]
	\caption{Comparison with Non-invasive Works}
	\label{table_example}
	\centering
	\begin{tabular}{lllllllll}
		\hline
		\textbf{Works}&\textbf{R}&\textbf{mARD}&\textbf{AvgE}&\textbf{MAD}&RMSE&Used&Measurement&Device\\
		\textbf{}&value&(\%)&(\%)&(mg/dl)&(mg/dl) &model&sample&cost\\
		\hline		 \hline
		Singh, et al. \cite{8727488}&0.80&-&-&-&-&Human&Saliva&Cheaper\\
		\hline
		Song, et al.	\cite{Song2015}&-&8.3&19&-&-&Human&Blood&Cheaper\\
		\hline
		Pai, et al.	\cite{Pai2018}&-&7.01&-&5.23&7.64&in-vitro&Blood&Costly\\
		\hline
		Dai, et al.	\cite{Dai2009}&-&5.99&5.58&-&-&in-vivo&Blood&Cheaper\\
		\hline
		Beach, et al.	\cite{Beach2005}&-&-&7.33&-&-&in-vitro&Solution&-\\
		\hline
		Ali, et al.	\cite{Ali2017}&-&8.0&-&-&-&Human&Blood&Cheaper\\
		\hline
		Haxha, et al. \cite{haxha2016optical}&0.96&-&-&-&33.49&Human&Blood&Cheaper\\
		\hline
		Jain, et al. (iGLU) \cite{Jain_IEEE-MCE_2020-Jan_iGLU1}&0.95&6.65&7.30&12.67&21.95&Human&Blood&Cheaper\\
		\hline
		\textbf{Proposed Work (iGLU 2.0)}&\textbf{0.97}&\textbf{4.86}&\textbf{4.88}&\textbf{9.42}&\textbf{13.57}&\textbf\textbf{Human}&\textbf{Blood}&\textbf{Cheaper}\\
		\hline
	\end{tabular}
\end{table*}


\section{Conclusions and Future Directions of This Research}

Non-invasive serum glucose measurement device is proposed which is based on short-wave NIR spectroscopy with two specific wavelengths (940 and 1300 nm). The device is validated and tested through healthy, prediabetic and diabetic patients. The combination of reflection and absorption of NIR light using MPR model based calibration is implemented for non-invasive serum glucose measurement. The proposed device is the combination of optical detection and optimized machine earning regression model which is an innovative approach for precise serum glucose measurement. For device validation, all samples have been taken from persons aged 17-80. During device calibration for serum glucose, the capillary glucose has also been calculated from the same samples for error analysis. During statistical analysis, 0.79 coefficient of determination is calculated for serum glucose calibration using MPR based model with polynomial degree 3 which shows a better result compared to capillary glucose calibration. 
It has been observed that $AvgE$ and $mARD$ represent better results of calibration and validation for serum glucose compared to capillary glucose. After analysis of predicted serum glucose values, 100\% samples come in zone A.

In the future research, we will consider Internet-of-Medical-Things (IoMT) integration of the iGLU in healthcare CPS (H-CPS) so that the healthcare data can be securely stored in the cloud. We will also evaluate the options of edge-computing and IoT-computing paradigms for fast response as well as long-term storage of the user's records. This can be have significant impact on smart living component of smart healthcare for smart homes \cite{Mohanty_IEEE-MCE_2020-Jan_Editorial}. The long-term goal of iGLU project is an unified H-CPS for non-invasive glucose level monitoring and release of appropriate level of insulin with full-proof security arrangement \cite{Jain_WF-IoT-2020_Insulin-Delivery-System}.

\section*{Acknowledgment}

The authors would like to thank Dispensary and System Level Design and Calibration Testing Lab, Malaviya National of Technology, Jaipur, Rajasthan (India) and special thanks for MHRD funded SMDP Lab which has provided the support of hardware components for the experimental implementation. Data collection has been done under inspection of him following medical protocols.

\bibliographystyle{IEEEtran}

\begin{thebibliography}{10}
\providecommand{\url}[1]{#1}
\csname url@samestyle\endcsname
\providecommand{\newblock}{\relax}
\providecommand{\bibinfo}[2]{#2}
\providecommand{\BIBentrySTDinterwordspacing}{\spaceskip=0pt\relax}
\providecommand{\BIBentryALTinterwordstretchfactor}{4}
\providecommand{\BIBentryALTinterwordspacing}{\spaceskip=\fontdimen2\font plus
\BIBentryALTinterwordstretchfactor\fontdimen3\font minus
  \fontdimen4\font\relax}
\providecommand{\BIBforeignlanguage}[2]{{%
\expandafter\ifx\csname l@#1\endcsname\relax
\typeout{** WARNING: IEEEtran.bst: No hyphenation pattern has been}%
\typeout{** loaded for the language `#1'. Using the pattern for}%
\typeout{** the default language instead.}%
\else
\language=\csname l@#1\endcsname
\fi
#2}}
\providecommand{\BIBdecl}{\relax}
\BIBdecl

\bibitem{Zhu_MCE_2019-Sep}
H.~{Zhu}, C.~K. {Wu}, C.~H. {KOO}, Y.~T. {Tsang}, Y.~{Liu}, H.~R. {Chi}, and
  K.~{Tsang}, ``{Smart Healthcare in the Era of Internet-of-Things},''
  \emph{IEEE Consumer Electronics Magazine}, vol.~8, no.~5, pp. 26--30,
  September 2019.

\bibitem{Mohanty_CEM_2016-Jul}
S.~P. {Mohanty}, U.~{Choppali}, and E.~{Kougianos}, ``{Everything you wanted to
  know about smart cities: The Internet of things is the backbone},''
  \emph{IEEE Consumer Electronics Magazine}, vol.~5, no.~3, pp. 60--70, July
  2016.

\bibitem{Mohanty_IEEE-MetroCon-2019_Invited-Talk}
\BIBentryALTinterwordspacing
\vspace{0mm}S. P.~Mohanty, ``{Smart Healthcare - Demystified},'' iEEE
  MetroCon2019 Invited TalkHurst Conference Center, TX, 06 Nov 2019. [Online].
  Available:
  \url{http://www.smohanty.org/Presentations/2019/Mohanty_IEEE-MetroCon-2019_Invited-Talk_Smart-Healthcare.pdf}
\BIBentrySTDinterwordspacing

\bibitem{Jain_IEEE-MCE_2020-Jan_iGLU1}
P.~Jain, A.~M. Joshi, and S.~P. Mohanty, ``{iGLU: An Intelligent Device for
  Accurate Non-Invasive Blood Glucose-Level Monitoring in Smart Healthcare},''
  \emph{IEEE Consumer Electronics Magazine}, vol.~9, no.~1, pp. 35--42, January
  2020.

\bibitem{Jain_arXiv_2019-Nov30-1911-04471_iGLU1}
\BIBentryALTinterwordspacing
\vspace{0mm}P. Jain, A.~M. Joshi, and S.~P. Mohanty, ``{iGLU 1.0: An Accurate
  Non-Invasive Near-Infrared Dual Short Wavelengths Spectroscopy based
  Glucometer for Smart Healthcare},'' \emph{arXiv Electrical Engineering and
  Systems Science}, vol. abs/1911.04471, 2019. [Online]. Available:
  \url{http://arxiv.org/abs/1911.04471}
\BIBentrySTDinterwordspacing

\bibitem{IEEE_Std_11073-10425-2017}
``{IEEE Health informatics--Personal health device communication - Part 10425:
  Device Specialization--Continuous Glucose Monitor (CGM)},'' \emph{IEEE Std
  11073-10425-2017 (Revision of IEEE Std 11073-10425-2014) - Redline}, pp.
  1--137, Jan 2018.

\bibitem{Wang_TBCS_2017-Oct}
G.~{Wang}, M.~D. {Poscente}, S.~S. {Park}, C.~N. {Andrews}, O.~{Yadid-Pecht},
  and M.~P. {Mintchev}, ``{Wearable Microsystem for Minimally Invasive,
  Pseudo-Continuous Blood Glucose Monitoring: The e-Mosquito},'' \emph{IEEE
  Transactions on Biomedical Circuits and Systems}, vol.~11, no.~5, pp.
  979--987, Oct 2017.

\bibitem{Beach2005}
R.~D. {Beach}, R.~W. {Conlan}, M.~C. {Godwin}, and F.~{Moussy}, ``Towards a
  miniature implantable in vivo telemetry monitoring system dynamically
  configurable as a potentiostat or galvanostat for two- and three-electrode
  biosensors,'' \emph{IEEE Transactions on Instrumentation and Measurement},
  vol.~54, no.~1, pp. 61--72, Feb 2005.

\bibitem{demitri2017measuring}
N.~Demitri and A.~M. Zoubir, ``Measuring blood glucose concentrations in
  photometric glucometers requiring very small sample volumes,'' \emph{IEEE
  Transactions on Biomedical Engineering}, vol.~64, no.~1, pp. 28--39, 2017.

\bibitem{acciaroli2018reduction}
G.~Acciaroli, M.~Vettoretti, A.~Facchinetti, G.~Sparacino, and C.~Cobelli,
  ``Reduction of blood glucose measurements to calibrate subcutaneous glucose
  sensors: A bayesian multiday framework,'' \emph{IEEE Transactions on
  Biomedical Engineering}, vol.~65, no.~3, pp. 587--595, 2018.

\bibitem{Pai2018}
P.~P. Pai, A.~De, and S.~Banerjee, ``Accuracy enhancement for noninvasive
  glucose estimation using dual-wavelength photoacoustic measurements and
  kernel-based calibration,'' \emph{IEEE Transactions on Instrumentation and
  Measurement}, vol.~67, no.~1, pp. 126--136, 2018.

\bibitem{Yin_2019}
\BIBentryALTinterwordspacing
H.~Yin, B.~Mukadam, X.~Dai, and N.~Jha, ``{DiabDeep: Pervasive Diabetes
  Diagnosis based on Wearable Medical Sensors and Efficient Neural Networks},''
  \emph{IEEE Transactions on Emerging Topics in Computing}, pp. 1--1, 2019.
  [Online]. Available: \url{http://dx.doi.org/10.1109/tetc.2019.2958946}
\BIBentrySTDinterwordspacing

\bibitem{8347021}
M.~S. {Prasad}, R.~{Chen}, Y.~{Li}, D.~{Rekha}, D.~{Li}, H.~{Ni}, and N.~Y.
  {Sreedhar}, ``Polypyrrole supported with copper nanoparticles modified alkali
  anodized steel electrode for probing of glucose in real samples,'' \emph{IEEE
  Sensors Journal}, vol.~18, no.~13, pp. 5203--5212, July 2018.

\bibitem{8727488}
A.~K. {Singh} and S.~K. {Jha}, ``Fabrication and validation of a handheld
  non-invasive, optical biosensor for self-monitoring of glucose using
  saliva,'' \emph{IEEE Sensors Journal}, vol.~19, no.~18, pp. 8332--8339, Sep.
  2019.

\bibitem{Dai2009}
T.~Dai and A.~Adler, ``In vivo blood characterization from bioimpedance
  spectroscopy of blood pooling,'' \emph{IEEE Transactions on Instrumentation
  and Measurement}, vol.~58, no.~11, p. 3831, 2009.

\bibitem{Song2015}
K.~Song, U.~Ha, S.~Park, J.~Bae, and H.~J. Yoo, ``An impedance and
  multi-wavelength near-infrared spectroscopy ic for non-invasive blood glucose
  estimation,'' \emph{IEEE Journal of Solid-State Circuits}, vol.~50, no.~4,
  pp. 1025--1037, April 2015.

\bibitem{de2016optical}
L.~R. De~Pretto, T.~M. Yoshimura, M.~S. Ribeiro, and A.~Z. de~Freitas,
  ``Optical coherence tomography for blood glucose monitoring in vitro through
  spatial and temporal approaches,'' \emph{Journal of Biomedical Optics},
  vol.~21, no.~8, p. 086007, 2016.

\bibitem{pirnstill2012vivo}
C.~W. Pirnstill, B.~H. Malik, V.~C. Gresham, and G.~L. Cot{\'e}, ``In vivo
  glucose monitoring using dual-wavelength polarimetry to overcome corneal
  birefringence in the presence of motion,'' \emph{Diabetes technology \&
  therapeutics}, vol.~14, no.~9, pp. 819--827, 2012.

\bibitem{shih2015noninvasive}
W.-C. Shih, K.~L. Bechtel, and M.~V. Rebec, ``Noninvasive glucose sensing by
  transcutaneous raman spectroscopy,'' \emph{Journal of biomedical optics},
  vol.~20, no.~5, p. 051036, 2015.

\bibitem{Fernandez2020}
\BIBentryALTinterwordspacing
C.~Fernandez, ``{Needle-Free Diabetes Care: 8 Devices that Painlessly Measure
  Blood Glucose},'' 2020. [Online]. Available:
  \url{https://www.labiotech.eu/diabetes/needle-free-glucose-monitoring-for-diabetes-medtech/}
\BIBentrySTDinterwordspacing

\bibitem{2020me}
\BIBentryALTinterwordspacing
``Non-invasive blood glucose meter,'' last visited 24 Jan 2020. [Online].
  Available:
  \url{https://www.2mel.nl/project/non-invasive-blood-glucose-meter/}
\BIBentrySTDinterwordspacing

\bibitem{sugarbeat}
\BIBentryALTinterwordspacing
``{SugarBEAT: "The World’s First Non-Invasive Glucose Monitor"},'' last
  visited 24 Jan 2020. [Online]. Available:
  \url{https://www.healthline.com/diabetesmine/non-invasive-sugarbeat-cgm-diabetes#1}
\BIBentrySTDinterwordspacing

\bibitem{omelon}
\BIBentryALTinterwordspacing
``Omelon b2 non invasive glucometer,'' last visited 24 Jan 2020. [Online].
  Available:
  \url{https://diabetestalk.net/blood-sugar/omelon-b2-non-invasive-glucometer}
\BIBentrySTDinterwordspacing

\bibitem{monte2011non}
E.~Monte-Moreno, ``Non-invasive estimate of blood glucose and blood pressure
  from a photoplethysmograph by means of machine learning techniques,''
  \emph{Artificial intelligence in medicine}, vol.~53, no.~2, pp. 127--138,
  2011.

\bibitem{habbu2019estimation}
S.~Habbu, M.~Dale, and R.~Ghongade, ``Estimation of blood glucose by
  non-invasive method using photoplethysmography,'' \emph{S{\=a}dhan{\=a}},
  vol.~44, no.~6, p. 135, 2019.

\bibitem{sharma2013efficient}
S.~Sharma, M.~Goodarzi, L.~Wynants, H.~Ramon, and W.~Saeys, ``Efficient use of
  pure component and interferent spectra in multivariate calibration,''
  \emph{Analytica chimica acta}, vol. 778, pp. 15--23, 2013.

\bibitem{Ali2017}
H.~{Ali}, F.~{Bensaali}, and F.~{Jaber}, ``Novel approach to non-invasive blood
  glucose monitoring based on transmittance and refraction of visible laser
  light,'' \emph{IEEE Access}, vol.~5, pp. 9163--9174, 2017.

\bibitem{haxha2016optical}
S.~Haxha and J.~Jhoja, ``Optical based noninvasive glucose monitoring sensor
  prototype,'' \emph{IEEE Photonics Journal}, vol.~8, no.~6, pp. 1--11, 2016.

\bibitem{jain2019precise}
P.~Jain, R.~Maddila, and A.~M. Joshi, ``{A precise non-invasive blood glucose
  measurement system using NIR spectroscopy and Huber’s regression model},''
  \emph{Optical and Quantum Electronics}, vol.~51, no.~2, p.~51, 2019.

\bibitem{Quesada_2020_5_algorithms_to_train_a_neural_network}
\BIBentryALTinterwordspacing
``Five algorithms to train a neural network,'' last visited 24 Jan 2020.
  [Online]. Available:
  \url{https://www.neuraldesigner.com/blog/5_algorithms_to_train_a_neural_network}
\BIBentrySTDinterwordspacing

\bibitem{seo2009clinical}
J.~A. Seo, N.~H. Kim, S.~G. Yun, C.~H. Cho, J.~H. Yang, C.~S. Lim, Y.~K. Kim,
  and K.~N. Lee, ``Clinical evaluation of sd check gold as point-of-care
  glucose meter,'' \emph{J Lab Med Qual Assur}, vol.~31, no.~2, p. 261, 2009.

\bibitem{Clarke2005}
W.~L. Clarke, ``{The original Clarke error grid analysis (EGA)},''
  \emph{Diabetes Technology \& Therapeutics}, vol.~7, no.~5, pp. 776--779,
  2005.

\bibitem{Mohanty_IEEE-MCE_2020-Jan_Editorial}
S.~P. {Mohanty}, ``{Consumer Technologies Build Smart Homes},'' \emph{IEEE
  Consumer Electronics Magazine}, vol.~9, no.~1, pp. 4--5, Jan 2020.

\bibitem{Jain_WF-IoT-2020_Insulin-Delivery-System}
P.~Jain, A.~M. Joshi, and S.~P. Mohanty, ``{iGLU 1.1: Towards a Glucose-Insulin
  Model based Closed Loop IoMT Framework for Automatic Insulin Control of
  Diabetic Patients},'' in \emph{Proceedings of the IEEE 6th World Forum on
  Internet of Things (WF-IoT)}, 2020, p. Under Review.

\end{thebibliography}


\pagebreak

\section*{Authors' Biographies}


\begin{minipage}[htbp]{\columnwidth}
\begin{wrapfigure}{l}{1.0in}
	\vspace{-0.4cm}
\includegraphics[width=1.0in,keepaspectratio]{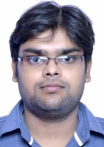}
	\vspace{-0.3cm}
\end{wrapfigure}
\noindent
\textbf{Prateek Jain} (GS'18) earned his B.E. degree in Electronics Engineering from Jiwaji University, India in 2010 and Master degree from ITM University Gwalior. He is a Research Scholar at th ECE department of MNIT, Jaipur. His current research interest includes VLSI design, Biomedical Systems and Instrumentation. He is an author of 14 peer-reviewed publications. He is a regular reviewer of 12 journals and 10 conferences.
\end{minipage}

\vspace{1.8cm}

\begin{minipage}[htbp]{\columnwidth}
\begin{wrapfigure}{l}{1.0in}
\vspace{-0.4cm}
\includegraphics[width=1.0in,keepaspectratio]{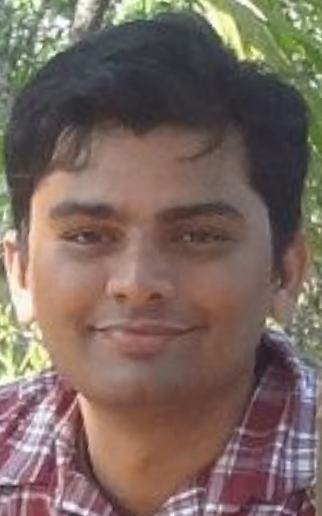}
	\vspace{-0.5cm}
\end{wrapfigure}
\noindent
\textbf{Amit M. Joshi} (M'08) has completed his M.Tech (by research) in 2009 and obtained Doctoral of Philosophy degree (Ph.D) from National Institute of Technology, Surat in August 2015. He is currently an Assistant Professor at National Institute of Technology, Jaipur since July 2013. His area of specialization is Biomedical signal processing, Smart healthcare, VLSI DSP Systems and embedded system design. He has published six book chapters and also published 50+ research articles in peer reviewed international journals/conferences. He has served as a reviewer of technical journals such as IEEE Transactions, Springer, Elsevier and also served as Technical Programme Committee member for IEEE conferences. He also received UGC Travel fellowship, SERB DST Travel grant  and CSIR Travel fellowship to attend IEEE Conferences in VLSI and Embedded System. He has served session chair at various IEEE Conferences like TENCON -2016, iSES-2018, ICCIC-14. He has already supervised 18 M.Tech projects and 14 B.Tech projects in the field of VLSI and Embedded Systems and VLSI DSP systems. He is currently supervising six  Ph.D. students.
\end{minipage}

\vspace{0.5cm}

\begin{minipage}[htbp]{\columnwidth}
	\begin{wrapfigure}{l}{1.00in}
		\vspace{-0.3cm}
		\includegraphics[width=1.0in,keepaspectratio]{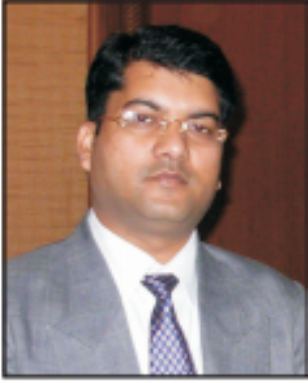}
		\vspace{-0.5cm}
	\end{wrapfigure}
\noindent
\textbf{Navneet Agrawal} is an MD, PhD, FCCP, FIACM, Diploma in Diabetology Consultant  Diabetologist and Director, Diabetes, Obesity and Thyroid Centre, Gwalior. He is Director of Academy of Medical and Allied Sciences, offering courses for medical and paramedical students. He is Regional Faculty of Public Health Foundation India  for course CCEBDM and CCGDM and regional faculty for IDF National diabetes Educator Programme. He is an Editor of various journals including: Annals of Diabetes and Clinical Complications (ADCC), Journal of Indian Academy of Clinical Medicine,  Electronic Physician, South Asia, International Journal of Clinical Cases and Investigation, International Journal of Pharmaceutical Sciences Review and Research, 
Journal of Medical Nutrition and Nutraceuticals, IP Journal of Nutrition, Metabolism and Health Science.  He has (co)authored more than 80 articles. 
He is awarded as Times of India Health Icon 2017, Distinguished Diabetologist 2016, Best Youth Doctor Gwalior 2016, Diabetes India Awareness Initiative Award  2015, and Agravansh Ratn 2012.
Outstanding Young person of Gwalior 2010, Chikitsa  Ratn 2010, International Travel Grant Fellowship 2009. 
\end{minipage}

\vspace{0.5cm}

\begin{minipage}[htbp]{\columnwidth}
	\begin{wrapfigure}{l}{1.00in}
		\vspace{-0.3cm}
		\includegraphics[width=1.0in,keepaspectratio]{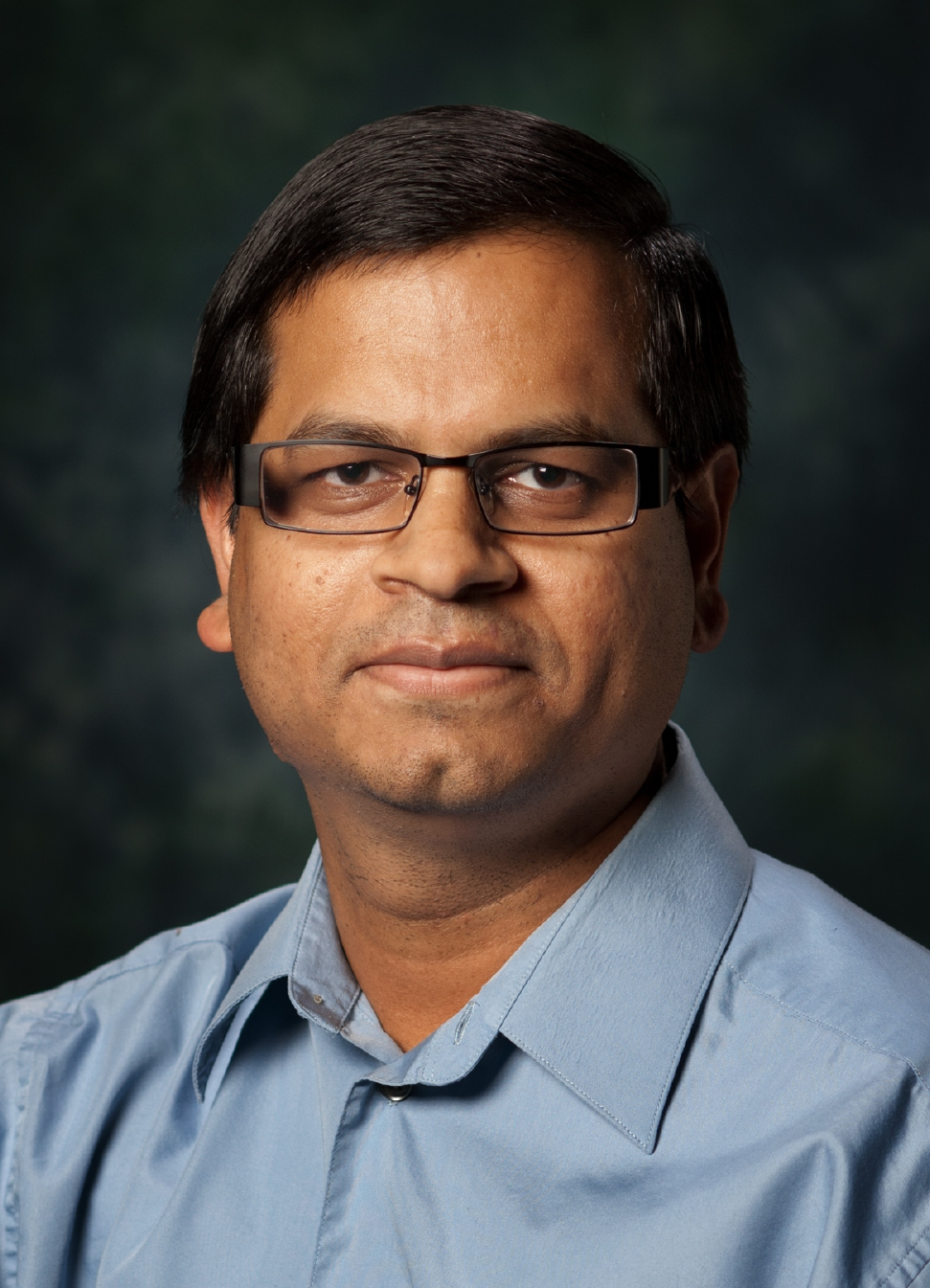}
		\vspace{-0.5cm}
	\end{wrapfigure}
	\noindent
\textbf{Saraju P. Mohanty} (SM'08) received the bachelor's degree (Honors) in electrical engineering from the Orissa University of Agriculture and Technology, Bhubaneswar, in 1995, the master's degree in Systems Science and Automation from the Indian Institute of Science, Bengaluru, in 1999, and the Ph.D. degree in Computer Science and Engineering from the University of South Florida, Tampa, in 2003. He is a Professor with the University of North Texas. His research is in ``Smart Electronic Systems'' which has been funded by National Science Foundations (NSF), Semiconductor Research Corporation (SRC), U.S. Air Force, IUSSTF, and Mission Innovation Global Alliance. He has authored 300 research articles, 4 books, and invented 4 U.S. patents. His has Google Scholar citations with an h-index of 34 and i10-index of 125, with 5000+ citations. He was a recipient of 11 best paper awards, IEEE Consumer Electronics Society Outstanding Service Award in 2020 for contributions to the IEEE Consumer Electronics society, the IEEE-CS-TCVLSI Distinguished Leadership Award in 2018 for services to the IEEE and to the VLSI research community, and the 2016 PROSE Award for Best Textbook in Physical Sciences and Mathematics category from the Association of American Publishers for his Mixed-Signal System Design book published by McGraw-Hill. He has delivered 9 keynotes and served on 5 panels at various International Conferences. He has been serving on the editorial board of several peer-reviewed international journals, including IEEE Transactions on Consumer Electronics, and IEEE Transactions on Big Data. 
He is currently the Editor-in-Chief (EiC) of the IEEE Consumer Electronics Magazine (MCE). 
\end{minipage}

\end{document}